\documentclass[conference]{IEEEtran}
\usepackage{caption}
\usepackage{algorithm}
\usepackage{algorithmic}
\usepackage{graphicx}
\usepackage{enumerate}
\usepackage{amsthm}
\usepackage{epstopdf}
\usepackage{amsmath,amssymb,epsfig,graphicx,slidesec,colortbl}
\usepackage[absolute,overlay]{textpos}
\usepackage{multimedia,algorithm,algorithmic,multirow,url}
\usepackage{cite}
\usepackage{subfig}
\usepackage{color}
\usepackage{url}
\usepackage{enumitem}

\newtheorem{assumption}{Assumption}
\newtheorem{game}{Game}
\newtheorem{definition}{Definition}
\newtheorem{theorem}{Theorem}

\newtheorem{observation}{Observation}
\DeclareMathOperator*{\argmax}{arg\,max}

\usepackage{geometry}
\IEEEoverridecommandlockouts
\begin{document}

\title{Incentivizing Data Contribution in \\Cross-Silo Federated Learning}
\author{Chao Huang, Shuqi Ke, Charles Kamhoua, Prasant Mohapatra, and Xin Liu

}

% use for special paper notices
%\IEEEspecialpapernotice{(Invited Paper)}

% make the title area
\maketitle

% As a general rule, do not put math, special symbols or citations
% in the abstract

\begin{abstract}
In cross-silo federated learning, clients (e.g., organizations) train a shared global model using local data. However, due to
% business competition and  
privacy concerns, the clients may not contribute enough data points during training. 
To address this issue, we propose a general incentive framework where the profit/benefit obtained from the global model can be appropriately allocated to clients to incentivize data contribution. We formulate the clients' interactions as a data contribution game and study its equilibrium. We characterize conditions for an equilibrium to exist, and prove that each client's equilibrium data contribution increases in its data quality and decreases in the privacy sensitivity. We further conduct experiments using CIFAR-10 and show that the results are consistent with the analysis. Moreover, we show that practical allocation mechanisms such as linearly proportional, leave-one-out, and Shapley-value incentivize more data contribution from clients with higher-quality data, in which leave-one-out tends to achieve the highest global model accuracy at equilibrium.
\end{abstract}

\section{Introduction}\label{introduction}
%\subsection{Motivations}
Federated learning (FL) is a decentralized machine learning paradigm where multiple clients collaboratively train a global model under the orchestration of a central server \cite{MAL-083}. %\footnote{Part of the results are submitted to ICML 2022.}
%Clients can keep their data local and only upload model updates (e.g., represented by parameters or gradients) trained by their data. 
%More specifically, each client trains a local model using its own data during the training process and then uploads the model updates to the central server. The central server appropriately aggregates all clients' uploaded model updates to generate a global model during the aggregation process.  It then sends back the global model to clients for further training. The iteration between the training  and aggregation processes terminates until the global model converges.
%\footnote{We have released our codes for the numerical experiments at XXX.}
Based on the scale and participating clients, FL can be classified into two types: cross-device FL and cross-silo FL \cite{huang2022cross}. In cross-device FL, clients are usually small distributed entities (e.g., smartphones, wearables, and edge devices), with each having a relatively small amount of local data.
Hence, for cross-device FL to succeed, it requires many edge devices to participate in the training process. 
In cross-silo FL, however, clients are often companies or organizations (e.g., hospitals and banks). The number of participants is relatively small, and each client is expected to participate in the entire training process.

This paper focuses on cross-silo FL, and its practical examples abound. In the medical and healthcare domain, Owkin collaborates with pharmaceutical companies to train models for drug discovery based on highly sensitive screening datasets \cite{owkin}. FeatureCloud utilizes FL to perform biomedical data analysis (e.g., COVID) \cite{featurecloud}.  In the financial area, WeBank and Swiss Re cooperatively conduct data analysis and provide financial and insurance services \cite{SwissReWebank}. 

The success of a global model requires the clients to contribute sufficient data for local model training. 
%In cross-silo FL, however, the organizations can be not only collaborators but also business competitors. Clients may be self-interested and not use enough data for training, which jeopardizes the performance of the global model.  
In cross-silo FL, however, the training data can be highly-sensitive (e.g.,  medical records and financial data), and hence the clients may have substantial \textbf{privacy concerns}.  Even if FL does not directly expose clients' local data, recent research has shown that FL can still be vulnerable to various privacy attacks when communicating model parameters \cite{noble2022differentially,pmlr-v162-wen22a}. Such potential privacy risks may deter clients from using enough data for training. In fact, a recent study \cite{fraboni2021free} identified the free-rider phenomenon in FL, where clients do not contribute any data in model training but enjoy (i.e., free-ride on) the well-trained global model. These observations motivate the first key question in this paper:

\textbf{Key Question 1}: \textit{How to incentivize clients to contribute data for FL training when they have privacy concerns?}

To answer \textit{Key Question 1}, 
%In fact, recent work has studied the free rider phenomenon in FL \cite{fraboni2021free,huang2020exploratory}. For example, the work in \cite{fraboni2021free} establishes an approach to inspect the clients' distribution for the detection of free-rider attacks. Free riding is detrimental to the global model accuracy, leading to the failure of cross-silo FL. Furthermore, the detected free riding behavior can damage the willingness of other participating clients to cooperate in the future. We aim to analyze and address the free-riding issue in this paper. 
we propose a general incentive framework where the total profit/benefit obtained from the global model can be appropriately allocated to clients to incentivize data contribution. More specifically, after  FL terminates, the clients can use the global model to generate some profits/benefits, e.g.,  via selling it in a model trading market \cite{chen2019towards,lin2021demonstration}, or via selling model-related services to customers \cite{wu2022mars}.  Then, the clients appropriately distribute the total profits according to a predefined mechanism. Intuitively, 
%It is important to note that the central server does not need to decide the profit allocation mechanism. In practice, the organizations themselves can negotiate a profit allocation mechanism and enforce its implementation via a contract  \cite{yu2020fairness}. 
a plausible mechanism should provide more profits to those who contribute more high-quality data to FL.
To this end, we evaluate practical mechanisms that allocate profits based on the evaluation of clients' contributions to FL, and study their impact on clients' data contribution decisions at a game-theoretical equilibrium. The equilibrium outcomes shed insights into the design of a desirable mechanism that incentivizes data contribution in cross-silo FL.

Furthermore, in practical cross-silo FL, clients can be highly heterogeneous. First and foremost, different clients may have \textbf{heterogeneous data qualities} \cite{yang2022robust,xu2022fedcorr}. Take medical diagnosis as an example. A hospital with more advanced equipment and better-trained professionals will likely provide cancer diagnosis with higher accuracy. This leads to a training dataset with higher quality (i.e., fewer noisy/wrong labels). Second, companies of different scales are expected to differ in their data capacities (i.e.,  the available number of local data points), and privacy sensitivities (i.e., the degree that an organization is concerned about privacy leakage).
%For example, a large hospital in a metropolitan area is likely to own more data than a small hospital in a rural area. 
%Third, different organizations are likely to have different privacy sensitivities . For example, companies located in countries that mandate stricter privacy protection are expected to be more privacy-sensitive than those with less stringent requirements for privacy protection.
%hospitals with good reputations can be more privacy-sensitive in using their patients' medical records for model training. 
%This might not be true for those hospitals which are more profit-driven and attach little importance to patient privacy. 
This motivates our second key question:

\textbf{Key Question 2}: \textit{How will client heterogeneity (w.r.t. data quality, data capacity, and privacy sensitivity) affect their data contribution decisions and  the mechanism design?}

To answer \textit{Key Question 2}, we proceed from both theoretical and empirical perspectives. Theoretically speaking, we first study the existence of an equilibrium and then investigate how the game equilibrium depends on the three-dimensional client heterogeneity. Empirically speaking, we conduct numerical experiments based on CIFAR-10  to validate our theoretical analysis and draw insights on the mechanism design.

%perform the game analysis considering two cases: the homogeneous case where clients have the same data capacity and privacy sensitivity; the heterogeneous case where clients can have different data capacities and privacy sensitivities. The game solutions under the two cases help us understand the impact of client heterogeneity and further enable us to better design the profit allocation mechanism for practical applications where cross-silo FL clients are heterogeneous.

%\subsection{Key Contributions}
The key contributions of this paper are listed as follows:
\begin{itemize}
	\item We propose a general incentive framework to incentivize data contribution in cross-silo FL. Specifically, the framework appropriately allocates  profits to clients based on clients' contributions to FL (Sec. \ref{system_model}).
	\item We formulate the heterogeneous clients' interactions as a data contribution game and analyze its equilibrium. Under mild conditions, we prove that an equilibrium exists. Then, we show that each client's equilibrium data contribution increases in its data quality and data capacity, but decreases in the privacy sensitivity. We further devise a best response algorithm to compute the equilibrium (Sec. \ref{game-formulation-sec}).
	%\item We devise a best response algorithm for the clients to compute the Nash equilibrium. We further customize four popular allocation mechanisms to our setting with heterogeneous data qualities (Sec. \ref{best-response}).  
	%We further evaluate four popular allocation mechanisms (all of which can be characterized by our framework) and investigate how they affect  
	%Under mild assumptions, we prove that the best response algorithm converges in polynomial time. 
	\item We conduct numerical experiments using CIFAR-10 and show that the results are consistent with our theoretical analysis. Moreover, we evaluate several popular allocation mechanisms, such as linearly proportional \cite{yu2020fairness}, leave-one-out \cite{wang2020principled}, and Shapley value \cite{jia2019towards}. We show that these mechanisms incentivize high-quality data contribution from clients, in which leave-one-out tends to induce the highest global model accuracy at equilibrium  (Sec. \ref{numerical}).
\end{itemize}

We organize the remainder of this paper as follows. We review related work in Section \ref{related_work}. We introduce the system model in Section \ref{system_model}. We present the game formulation and provide theoretical analysis in Section \ref{game-formulation-sec}. 
%We analyze the homogeneous client case and the heterogeneous client case in Sections \ref{equilibrium_analysis} and \ref{Heterogeneous_analysis}, respectively.
We show numerical results in Section \ref{numerical}.
% and conclude in Section \ref{conclusions}. 
We discuss challenges and opportunities in Section \ref{discussions} and conclude in Section \ref{conclusions}.
\section{Related Work}\label{related_work}

%Our work aims to address the free rider issue in cross-silo FL via proper profit allocation. Hence, we review the related work from three perspectives: cross-silo FL, free rider issue, and profit allocation.
\textbf{Data Quality in FL}:
Data quality is critical in FL, and there are two major metrics to quantify data quality: label imbalance and label noise. Existing studies focus on the label imbalance issue \cite{achituve2021personalized,huang2021personalized,li2021model,zhang2022personalized,hanzely2020lower}. For example, Arhituve et al. \cite{achituve2021personalized} proposed personalization approaches to addressing the label imbalance issue. Li at al. \cite{li2021model} used contrastive learning to improve local training.

Our work focuses on the under-explored but equally important issue of label noise in FL. There are some recent relevant studies, e.g., Yang et al. \cite{yang2022robust} proposed robust aggregation to mitigate label noise. Xu et al. \cite{xu2022fedcorr} described how to detect noisy labels. Our work takes a novel game-theoretical perspective on how label noise affects FL clients' strategic data contribution behaviors.

\textbf{Incentives in FL}: In practical FL systems where clients are self-interested and non-altruistic, they may be unwilling to participate/contribute enough data for model training without adequate incentives. Donahue et al. \cite{donahue2021model,donahue2021optimality} characterized conditions when clients participate in FL and form stable coalitions. Sim et al. \cite{sim2020collaborative} and Xu et al. \cite{xu2021gradient} proposed Shapley-value-based incentive schemes to enhance client collaboration. Fraboni et al. \cite{fraboni2021free} studied the free-riders' impact on FL, and Zhang et al. \cite{zhang2022enabling} devised a repeated game incentive solution to mitigate the free-rider issue.

Perhaps the most relevant work is \cite{blum2021one}, where authors investigated how clients decide data sampling strategies to satisfy the model accuracy requirement. However, \cite{blum2021one} studied two equilibrium concepts (i.e., stable equilibrium and envy-free equilibrium) that are potentially weaker than our considered Nash equilibrium \cite{osborne1994course}. Further, \cite{blum2021one} did not consider the incentive design or incorporate the important feature of data quality (e.g., label noise).
%While most existing studies focus on cross-device FL, some recent works analyzed cross-silo FL.   The work in \cite{marfoq2020throughput} studies the topology design to maximize the number of communication rounds per unit time. The work in \cite{heikkila2020differentially} combines additively homomorphic secure summation protocols with differential privacy to preserve privacy. The study in \cite{huang2021personalized} proposes a heuristic approach to address the non-i.i.d. issue. The study in \cite{li2020practical} devises an algorithm FedKT that enables few-shot cross-silo FL. However, none of these studies analyzes how to address the free-rider issues. 

%There are some existing studies on identifying the free-rider issue in FL \cite{lin2019free,huang2020exploratory,fraboni2021free}. For example, the work in \cite{fraboni2021free} establishes an approach to inspect the clients' distribution for detecting the free-rider attacks. However, these papers focus on identifying the free rider phenomenon instead of analyzing its motivation or addressing the issue effectively. The work in \cite{zhang2021enabling} proposed a repeated game solution to address this issue. However, their solution requires the clients to participate in a game with infinite phases. Our work builds upon free rider detection and seeks to mitigate the issue within one phase. 

\textbf{Data Valuation in FL}: Our incentive framework is inspired by data valuation in FL. 
Existing studies along this line focus on either fairness or efficiency (i.e., fast computation) in data valuation \cite{ghorbani2020distributional,jia2019towards,song2019profit,ghorbani2019data,lyu2020collaborative}. However, they did not consider the strategicness of clients or the need for equilibrium. Our work studies how data valuation (which corresponds to the profit allocation mechanism) affects the clients' strategic data contribution decisions.     
%Profit allocation in FL pertains to compensating the clients for contributing their quality data for model training.  To achieve this, the central server needs to effectively and fairly evaluate the clients' contribution. 
%Existing studies mainly focus on three contribution evaluation/profit allocation mechanisms: linear proportional \cite{zhan2020learning}, leave-one-out \cite{yu2020fairness,bhagoji2019analyzing}, and Shapley-value-based methods \cite{song2019profit,fan2021improving,ghorbani2019data}. However, these papers focus on deriving efficient and fair algorithms for contribution valuation/profit allocation, rather than analyzing their impact on the clients' free rider behaviors. Furthermore, existing studies do not consider the exponential mechanism for profit allocation in FL. We fill this gap by proposing an exponential mechanism that outperforms existing methods in incentivizing clients' data contribution.

In summary, we propose the first general incentive framework based on data valuation to incentivize data contribution in cross-silo FL with noisy labels.

\section{System Model}\label{system_model}
We first introduce a cross-silo FL process with noisy labels in Section \ref{cross-silo-model}, and then present the incentive framework in Section \ref{incentive-framework}. We model the clients' decisions and payoff functions in Section \ref{client-decisions}. 
\subsection{A Cross-Silo FL Process with Noisy Labels}\label{cross-silo-model}
We consider a set $\mathcal{N}=\{1,2,\cdots, N\}$ of clients (e.g., hospitals) who train a shared global model for an image classification problem with $C$ classes. Each client $n\in \mathcal{N}$ owns a private local dataset $\mathcal{D}_n=\{(\boldsymbol{x}_n^i, y_n^i)\}_{i=1}^{D_n}$ with $D_n \triangleq |\mathcal{D}_n|$.
In practice, different clients tend to have data with varying qualities, e.g.,  different label noise patterns  \cite{zhu2021federated}. Label noise refers to the mismatch between a ground truth label $y^*$ (e.g., provided by expert annotators) and a different observed label $y$ in a given dataset. Mathematically, label noise can be modeled as a label flipping process, in which each label in class $k\in C$ is mislabeled as class $l\neq k\in C$ with probability $p(y=l|y^*=k)$. Let $\epsilon_n$  denote the label noise/error rate of client $n$, where $\epsilon_n=p(y^i_n\neq k|{y_n^{i}}^*=k)$,  $\forall i$.
%More specifically, 
%suppose that there is an underlying \textit{correct} label distribution $P_{\pi}(y^*)$ (e.g., provided by expert annotators). Each client $n$'s data is generated from a possibly different/noisy distribution $P_n(y|\boldsymbol{x})$. Let $\epsilon_n=\sum_{c=1}^{C}|P_{\pi}(y=c|\boldsymbol{x})-P_{n}(y=c|\boldsymbol{x})|$ denote the proportion of noisy/erroneous labels of client $n$ and $\boldsymbol{\epsilon}\triangleq \{\epsilon_n\}_{n \in \mathcal{N}}$. 
A smaller $\epsilon_n$ implies that client $n$ has data of higher quality (i.e., fewer noisy labels). We denote  $\boldsymbol{\epsilon}\triangleq \{\epsilon_n\}_{n \in \mathcal{N}}$, and $\epsilon_n$ may vary across clients.
%\footnote{This paper focuses on label noise. In our supplementary material, we further investigate how label imbalance affects the results.}

%In this paper, we consider that different clients may have different data qualities, inspired by practical observations where clients' data labels are error prone \cite{}.   Specifically, let $\epsilon_n\in [0,1]$ denote the proportion of erroneous labels of client $n$, and $\epsilon_n$ can vary across clients (e.g., non-i.i.d.).

%characterized by $\epsilon_n \in [0,1]$. 
%In this paper, we assume that the data are i.i.d. across all clients \cite{zhang2021enabling,ding2020optimal}. This is reasonable in cases where different hospitals train a diagnosis model for the same diseases. Considering i.i.d. data also serves as a starting point and enables closed-form insights that might go beyond the i.i.d. case. We leave the study of non-i.i.d. data into future work.

Due to privacy concerns of data leakage, clients may not use all local data $\mathcal{D}_n$ for FL. We consider that
each client $n$ chooses a subset $\mathcal{S}_n \in \mathcal{D}_n$ with $s_n\triangleq|\mathcal{S}_n|$ for  the entire FL process \cite{blum2021one,zhang2022enabling} and $\boldsymbol{s}\triangleq\{s_n\}_{n\in \mathcal{N}}$.  
%Let $\boldsymbol{s}_n^i$ denote the $i$-th data sample in $\mathcal{S}_n$.
In cross-silo FL, the clients train a global model represented by $\boldsymbol{\mathcal{\omega}}$. Let $l(\boldsymbol{\mathcal{\omega}}; (\boldsymbol{x}_n^i, y_n^i))$ denote the loss function for each $(\boldsymbol{x}_n^i, y_n^i)\in \mathcal{S}_n$ under $\boldsymbol{\mathcal{\omega}}$, and $L_n(\boldsymbol{\mathcal{\omega}})$ denotes client $n$'s  expected local loss function averaged over  $l(\boldsymbol{\mathcal{\omega}}; (\boldsymbol{x}_n^i, y_n^i))$ for all data points in $\mathcal{S}_n$. The clients derive the optimal weights $\boldsymbol{\mathcal{\omega}^*}$ by minimizing the global loss function $	L(\boldsymbol{\mathcal{\omega}}) \triangleq  \sum_{n\in \mathcal{N}}\frac{s_n}{\sum_{n'\in\mathcal{N}}s_{n'} } L_n(\boldsymbol{\mathcal{\omega}})$.
%Define:
%\begin{itemize}
%	\item $l(\boldsymbol{\mathcal{\omega}}; \boldsymbol{s}_n^i)$: the loss function for  $\boldsymbol{s}_n^i \in \mathcal{S}_n$ under $\boldsymbol{\mathcal{\omega}}$.
%	\item $L_n(\boldsymbol{\mathcal{\omega}})$: the expected local loss function of client $n$ 
%	averaged over $l(\boldsymbol{\mathcal{\omega}}; \boldsymbol{s}_n^i)$ for all $\boldsymbol{s}_n^i \in \mathcal{S}_n$, and it has the following form:
%	\begin{equation}
	%		L_n(\boldsymbol{\mathcal{\omega}}) = \frac{1}{x_n}\sum_{\boldsymbol{s}_n^i \in \mathcal{S}_n} l(\boldsymbol{\mathcal{\omega}}; \boldsymbol{s}_n^i).
	%	\end{equation}
%	\item $L(\boldsymbol{\mathcal{\omega}})$: the global loss function with the following form:
%	\begin{equation}\label{global-loss}
	%		L(\boldsymbol{\mathcal{\omega}}) = \sum_{n\in \mathcal{N}}\frac{x_n}{\sum_{n'\in\mathcal{N}}x_{n'} } L_n(\boldsymbol{\mathcal{\omega}}).
	%	\end{equation}
%\end{itemize}
%
%The clients seek to derive the optimal weights of the global model $\boldsymbol{\mathcal{\omega}^*}$ that minimize the global loss function in (\ref{global-loss}):
%\begin{equation}
%	\boldsymbol{\mathcal{\omega}^*} = \arg\min_{\boldsymbol{\mathcal{\omega}} }\left\{L(\boldsymbol{\mathcal{\omega}})\right\}.
%\end{equation}

To derive $\boldsymbol{\mathcal{\omega}^*}$, cross-silo FL proceeds in multiple rounds using, for example, FedAvg \cite{MAL-083}.
% In this paper, we consider that the server adopts the widely used FedAvg algorithm to find $\boldsymbol{\mathcal{\omega}^*}$ \cite{pillutla2019robust}.  
%TALK ABOUT HOW FEDAVG IS REASONABLE
%
%TALK ABOUT ALL CLIENTS PARTICIPATE
%
%TALK ABOUT SYNCHRONOUS UPDATE SCHEME
In each round $r$, the clients perform the following three steps. First, clients download the global model $\boldsymbol{\mathcal{\omega}^{r-1}}$ generated from the previous round. Second, clients perform local training with  $\boldsymbol{\mathcal{\omega}^{r-1}}$ over the chosen dataset via  mini-batch SGD \cite{li2014efficient}. Third, clients derive model updates $\boldsymbol{\mathcal{\omega}^{r}_n}$ and send them to the server for global aggregation, i.e., $	\boldsymbol{\mathcal{\omega}^{r}} = \sum_{n\in \mathcal{N}}\frac{s_n}{\sum_{n'\in\mathcal{N}}s_{n'} } \boldsymbol{\mathcal{\omega}^{r}_n}$.\footnote{In this paper, we assume that $s_n, \forall n$ is known to the server (and the clients). This is a widely adopted assumption in literature  (e.g., FedAvg \cite{MAL-083} and FedProx \cite{li2020federated}) where the server uses $s_n$ to weight the clients' model updates during global aggregation. The server can also announce this information to the clients \cite{zhang2022enabling}. We leave the case where $s_n$ is unknown to future work.}
The three-step iteration stops until the global model converges.

\subsection{A General Incentive Framework}\label{incentive-framework}
As mentioned, due to privacy concerns, cross-silo clients may not contribute sufficient data for FL. To mitigate this issue, we propose a framework where the benefits/profits obtained from the global model can be appropriately allocated to incentivize data contribution. 
%clients commit to a mechanism that appropriately distributes the profit generated from the global model. 
Specifically, let $A(\boldsymbol{s}, \boldsymbol{\epsilon})$ denote the global model accuracy, which depends on both the clients' data contribution  $\boldsymbol{s}$ and the clients' data quality $\boldsymbol{\epsilon}$. We assume that the clients can generate a total profit $\Pi(A(\boldsymbol{s}, \boldsymbol{\epsilon}))$ via selling model-related services (e.g., loan interest determination in digital banking \cite{yang2019federated}). The profit amount $\Pi(A(\boldsymbol{s}, \boldsymbol{\epsilon}))$ is non-decreasing in $A(\boldsymbol{s}, \boldsymbol{\epsilon})$, meaning that a better model leads to a larger profit. Next, each client obtains a proportion of the total profit based on their contribution to FL, which we explain below.
%\begin{equation}\label{profit-function}
%P(x_n, \boldsymbol{x}_{-n}) = P_{\rm base} - A(x_n, \boldsymbol{x}_{-n}).
%\end{equation}
%Here, $P_{\rm base}$ is a constant capturing the base profit and it can be estimated via market research approaches \cite{yu2020fairness}. The term $ A(x_n, \boldsymbol{x}_{-n})$ is the penalty term, meaning that a larger model accuracy loss corresponds to a smaller amount of total profit.\footnote{It is important to note that this paper considers a simple linear model that characterizes the relationship between the total profit $P(x_n, \boldsymbol{x}_{-n})$ and the global model accuracy loss $A(x_n, \boldsymbol{x}_{-n})$. Such a linear form enables derivation of clean insights and is widely adopted in literature \cite{jiao2020toward}. Our analysis can be extended to other models where the total profit is negatively correlated with the global model accuracy loss.}
%where $\beta$ is the coefficient capturing the sensitivity of the model profit $	P(x_n, \boldsymbol{x}_{-n})$ in the accuracy loss $A(x_n, \boldsymbol{x}_{-n})$. 

Consider the mapping $ \{\boldsymbol{s}, \boldsymbol{\epsilon}\} \mapsto \{I_n(\boldsymbol{s}, \boldsymbol{\epsilon})\}_{n\in \mathcal{N}}$ that 
%\begin{equation}
%\mathcal{F}: \{\boldsymbol{s}, \boldsymbol{\epsilon}\} \mapsto \{I_n(\boldsymbol{s}, \boldsymbol{\epsilon})\}_{n\in \mathcal{N}},
%\end{equation}
denotes a general profit allocation mechanism, where $I_n(\boldsymbol{s}, \boldsymbol{\epsilon})$ is the contribution index of client $n$. Each client $n$ obtains a proportion $g_n(\boldsymbol{s}, \boldsymbol{\epsilon})\triangleq \frac{I_n(\boldsymbol{s}, \boldsymbol{\epsilon})}{\sum_{n'=1}^{N}I_{n'}(\boldsymbol{s}, \boldsymbol{\epsilon})}$ of the total profit. The following examples are widely adopted in the literature for profit allocation:

\textbf{\underline{EG}alitarian (EG)} \cite{yu2020fairness}: clients equally share profit, i.e., 
\begin{equation}\label{eg}
I_n^{ EG}(\boldsymbol{s}, \boldsymbol{\epsilon})=1/N.
\end{equation}
%i.e., $I_n^{ EG}(\boldsymbol{s}, \boldsymbol{\epsilon})=1/N$, $\forall n$.\\
\textbf{\underline{L}inearly \underline{P}roportional (LP)} \cite{yu2020fairness}: a client's contribution index is linearly proportional to its data contribution, i.e.,  
\begin{equation}\label{lp}
I_n^{ LP}(\boldsymbol{s}, \boldsymbol{\epsilon})=(1-\epsilon_n)s_n.
\end{equation}
\textbf{\underline{L}eave-\underline{O}ne-\underline{O}ut (LOO)}  \cite{wang2020principled}: Each client $n$' contribution index is calculated by its marginal contribution to the global model accuracy, i.e., 
\begin{equation}\label{loo_ci}
\begin{aligned}
	I^{LOO}_n(\boldsymbol{s}, \boldsymbol{\epsilon}) =
	A(\{s_i\}_{i\in \mathcal{N}}, \boldsymbol{\epsilon})-A(\{s_j\}_{j \in \mathcal{N} \setminus \{n\}}, \boldsymbol{\epsilon}).
\end{aligned}
\end{equation}
\textbf{\underline{S}hapley \underline{V}alue (SV)} \cite{jia2019towards}:
Shapley value is a celebrated concept in cooperative game theory. It assigns a unique proportion of the total profit via the \textit{average} marginal contribution made by each client, i.e., 
\begin{equation}\label{sv}
\begin{aligned}
	I^{SV}_n(\boldsymbol{s}, \boldsymbol{\epsilon}) =\sum_{\mathcal{S} \subseteq \mathcal{N}\setminus\{n\}}\frac{A(\{s_i\}_{i\in \mathcal{S}\cup\{n\}}, \boldsymbol{\epsilon})\hspace{-1mm}-\hspace{-1mm}A(\{s_j\}_{j \in \mathcal{S} }, \boldsymbol{\epsilon})}{N{N-1 \choose |\mathcal{S}|}}.
\end{aligned}
\end{equation}
%\begin{Exam}{(\underline{L}eave-\underline{O}ne-\underline{O}ut (LOO) \cite{wang2020principled})}\label{loo} Each client $n$' contribution index is calculated by its marginal contribution to the global model accuracy, i.e., 
%	\begin{equation}\label{loo_ci}
%		\begin{aligned}
	%			I^{LOO}_n(\boldsymbol{s}, \boldsymbol{\epsilon}) =
	%			A(\{s_j\}_{j\in \mathcal{N}}, \boldsymbol{\epsilon})-A(\{s_j\}_{j \in \mathcal{N} \setminus \{n\}}, \boldsymbol{\epsilon}).
	%		\end{aligned}
%	\end{equation}
%	\end{Exam}
% i.e.,  
%\begin{equation}
%g_n(\boldsymbol{s}, \boldsymbol{\epsilon})=\frac{I_n(\boldsymbol{s}, \boldsymbol{\epsilon})}{\sum_{n'=1}^{N}I_{n'}(\boldsymbol{s}, \boldsymbol{\epsilon})}.
%\end{equation} 
%The clients compete with each other in terms of obtaining the total profit.

Note that 1) the implementation of LP requires the knowledge of $\epsilon_n$, but EG, LOO, and SV do not need such information; 2) while our experiments (Sec. \ref{numerical}) use the above mechanisms, our theoretical analysis (Sec. \ref{game-formulation-sec}) is based on the general framework that goes beyond these mechanisms. 
%In practice, the clients can negotiate a mechanism  themselves (e.g., via Nash bargaining \cite{zhang2022equality}).
\subsection{Clients' Strategy and Payoff}\label{client-decisions}
%In this subsection, we define each client's strategy and payoff function.
\textbf{Client Data Contribution Strategy}:
Each client $n$ randomly chooses a subset of local data $\mathcal{S}_n\in \mathcal{D}_n$ for local training throughout the entire FL process \cite{blum2021one},
%\footnote{Since data are i.i.d., it suffices to consider client $n$'s chosen data size $x_n$ when evaluating its impact on the global model.}
and we denote $s_n=|\mathcal{S}_n|\in [0, D_n]$ as client $n$'s data contribution level.

\textbf{Allocated Profit}: Based on the incentive framework, each client $n$ obtains an allocated profit: $g_n(\boldsymbol{s}, \boldsymbol{\epsilon})\cdot\Pi(A(\boldsymbol{s}, \boldsymbol{\epsilon}))$.

%In the following, we use two instances to exemplify the above mapping.  
%\begin{example}{(Uniform Mechanism)}
% Let $CI^{UR}_n(x_n, \boldsymbol{x}_{-n})$ denote the contribution index for client $n\in\mathcal{N}$ under the uniform mechanism. Then, we have
% \begin{equation}
% 	CI^{UR}_n (x_n, \boldsymbol{x}_{-n}) = \frac{1}{N}, \quad \forall n \in \mathcal{N}.
% 	\end{equation}
%	\end{example}
%
%\begin{example}{(Linearly Proportional)}
%	Let $CI^{LP}_n(x_n, \boldsymbol{x}_{-n})$ denote the contribution index for client $n\in\mathcal{N}$ under the linearly proportional mechanism. Then, we have
%	\begin{equation}
%		CI^{LP}_n(x_n, \boldsymbol{x}_{-n}) = \frac{x_n}{x_n + \sum_{n' \in \mathcal{N}\setminus\{n\}}x_{n'}}, \quad \forall n \in \mathcal{N}.
%	\end{equation}
%\end{example}

%\subsubsection{Costs}
%Clients participating in cross-silo FL incur
%three main types of costs: privacy cost, computation cost, and
%communication cost, which we elaborate on as follows:

\textbf{Privacy cost}:
%The clients in cross-silo federated learning incur privacy costs. For example, hospitals can be highly sensitive to their patients' personal data and hence are unwilling to contribute data for training. Notice that even if federated learning does not directly expose the detailed contents of the training data (whereas model updates are communicated), clients can still be vulnerable to various attacks such as the inference attack \cite{hu2021source}. In such cases, the clients' privacy can be compromised, which leads to privacy costs. 
Training with local data incurs privacy costs, especially for clients such as hospitals. For example, an adversary (e.g., a corrupted central server) can infer information
of clients' private training data through uploaded model
updates by launching inference attacks and model inversion
attacks \cite{pmlr-v162-wen22a}. Recent studies applied methods such as
differential privacy to mitigate this issue, but it causes model accuracy degradation. 
%they may not fully
%address the privacy concern. 
%In fact, cross-silo clients usually have stringent privacy requirements, e.g., hospitals can
%be highly sensitive to their patients’ medical data. 

We use
$C_n^{\rm pri}(s_n) = \mu_n\cdot  f(s_n)$ to denote client $n$'s privacy cost, where $\mu_n\ge 0$ represents a client's privacy sensitivity and $f(s_n)$
increases in $s_n$. Intuitively, if a privacy attack occurs, a client experiences more data leakage (i.e., larger privacy cost) when it uses more data for FL training.

%In this paper, we consider that each client $n$ incurs a linear privacy cost \cite{liao2020privacy}:
%\begin{equation}
%C_n^{pri}(x_n) = \mu_n x_n, \quad n\in \mathcal{N},
%\end{equation}
%where $\mu_n\ge 0$ presents the privacy sensitivity of clients.

\textbf{Computation and Communication Costs:}
%\subsubsection{Computation and Communication Costs}
Cross-silo FL also incurs computational costs (e.g., model training) and communication costs (e.g., uploading/downloading model updates).  However, cross-silo clients are companies or organizations who
are likely to have strong computational resources (e.g., powerful
servers) and reliable communication channels (e.g., high-speed
wired connections). Hence, we normalize the computation
and communication costs to be zero, as they are less major
concerns in cross-silo FL.\footnote{We concur that in cross-device FL, however, computation and communication costs can be the bottleneck since edge devices are usually resource-constrained \cite{gao2021convergence}.}
%The CPU energy consumption relies on various factors, such as the client's computing chip architecture, the number of CPU cycles to perform the local training, and the CPU processing speed (in cycles per second) \cite{tran2019federated}. Since the clients in cross-silo federated learning (e.g., companies or organizations) usually have strong computational resources, we consider that the computational cost for each client is a constant $C^{comp}_n \ge 0$.

%In a cross-silo federated learning process, clients upload their local model updates to the server for aggregation, and then download the updated global model for the next training rounds \cite{chang2020communication}. Since the cross-silo FL clients are expected to have reliable transmission networks (e.g.,  high-speed wired networks), we consider that the communication cost for each client is a constant $C^{comm}_n\ge 0$.

\textbf{Clients' Payoff Function}:
Each client $n$'s payoff function is defined as the difference between the allocated profit and the privacy cost:
\begin{equation}\label{payoff-function}
\begin{aligned}
U_n(\boldsymbol{s}, \boldsymbol{\epsilon})= g_n(\boldsymbol{s}, \boldsymbol{\epsilon})\cdot\Pi(A(\boldsymbol{s}, \boldsymbol{\epsilon})) - C_n^{\rm pri}(s_n).
\end{aligned}
\end{equation}
%The term $\frac{CI_n(x_n, \boldsymbol{x}_{-n})}{\sum_{n'\in\mathcal{N}}CI_{n'}(x_{n'}, \boldsymbol{x}_{-n'})}\cdot P(x_n, \boldsymbol{x}_{-n})$ captures the  profit allocated to client $n$. The terms $C_n^{\rm pri}(x_n)$, $C^{comp}_n$, and $C^{comm}_n$ represent client $n$'s privacy cost, computation cost, and communication cost, respectively.

We summarize the key notations in Table \ref{notation}, where the subscript $n$ corresponds to client $n$.
\begin{table}[t]
\caption{Key notations.} \label{notation}
\begin{center}
\begin{tabular}{l|l|l|l}
%	\textbf{Notation} 
\hline
&\textbf{Description} &
%\textbf{Notation} 
&\textbf{Description}\\
\hline
$s_n$ & data contribution &  $A(\cdot)$ & model accuracy\\
$\epsilon_n$  &label noise rate &$\Pi(\cdot)$  & profit function\\
$D_n$       &data capacity     & $I_n(\cdot)$& contribution index\\
$\mu_n$ &privacy sensitivity  &$g_n(\cdot)$& profit share \\
$C_n^{\rm pri}(\cdot)$& privacy cost & $U_n(\cdot)$ & payoff function\\
\hline
\end{tabular}
\end{center}
\vspace{-5mm}
\end{table}
%	$s_n$     &data contribution level of client $n$ &
%$\epsilon_n$       &label noise/error rate of client $n$\\
%$\mu_n$            &privacy sensitivity of client $n$
%&$D_n$       &data capacity of client $n$\\
%$A(\boldsymbol{s}, \boldsymbol{\epsilon})$ & global model accuracy
%&$\Pi(A(\boldsymbol{s}, \boldsymbol{\epsilon}))$  & profit generating function\\
%$I_n(\boldsymbol{s}, \boldsymbol{\epsilon})$ & contribution index of client $n$
%&$g_n(\boldsymbol{s}, \boldsymbol{\epsilon})$ & profit share of client $n$\\
%$C_n^{\rm pri}(s_n)$ & privacy cost of client $n$
%& $I_n(\boldsymbol{s}, \boldsymbol{\epsilon})$ & contribution index of client $n$\\
%$U_n(\boldsymbol{s}, \boldsymbol{\epsilon})$ & payoff function of client $n$

\section{Game Formulation and  Analysis}\label{game-formulation-sec}
We formulate the clients' data contribution game in 
Section \ref{formulation}. We analyze the equilibrium existence and properties in Sections \ref{equilibrium_existence} and \ref{equilibrium_properties}, respectively. We devise a best response algorithm to compute equilibrium in Section \ref{equilibrium_calculation}.

\subsection{Game Formulation}\label{formulation}
Each client strategically chooses its data contribution  $s_n$ to maximize its own payoff in (\ref{payoff-function}). Since each client's data contribution affects the global model accuracy and the profit allocated to other clients, the clients interact in a game-theoretical fashion. We model the interactions among the clients as a data contribution game below:
\begin{game}{(Data Contribution Game)}\label{game-formulation}
The data contribution game among the clients is defined as follows:
\begin{itemize}
\item Players: the set $\mathcal{N}$ of clients.
\item Strategies: each client $n$ decides the data contribution level $s_n\in [0, D_n]$ for local model training.
\item Payoffs: each client $n$ maximizes its own payoff in (\ref{payoff-function}).
\end{itemize}
\end{game}

%In Game \ref{game-formulation}, given others' decisions $\boldsymbol{x}_{-n}$, each client $n$ solves the problem below, which determines its best response.

%\begin{problem}{(Client $n$'s Best Response Problem)}\label{best-response}
%\begin{equation}
%\begin{aligned}
%	\max \ & \  u_n( x_n, \boldsymbol{x}_{-n})\\
%	\text{\rm{var.}} \ & \ x_n\in [0, D_n]. \\
%\end{aligned}
%\label{workerDecision}
%\end{equation}
%\end{problem}

We aim to solve the Nash equilibrium (NE) of Game \ref{game-formulation}, which is a stable outcome as no client can be better off via unilaterally changing its strategy.
\begin{definition}{(Nash Equilibrium)}  A profile $\boldsymbol{s}^*=\{s_n^\ast\}_{n \mathcal{N}}$ is a Nash equilibrium, if $\forall s_n' \in [0, D_n]$ and $\forall n \in \mathcal{N},$ 
\begin{equation}
U_n( s_n^\ast, \boldsymbol{s}_{-n}^\ast, \boldsymbol{\epsilon}) \ge 	U_n( s_n', \boldsymbol{s}_{-n}^\ast, \boldsymbol{\epsilon}), 
\end{equation}
where $\boldsymbol{s}_{-n}^*=\{s_j^*\}_{j \in\mathcal{N}\setminus \{n\}}$.
\end{definition}

At a Nash equilibrium, each client's strategy is a best response to the strategies of the other clients, i.e., the strategy profile is the fixed point of all clients' best responses. 
%Notice that a client $n$'s strategy choice is the optimal solution of Problem \ref{best-response}, and it is a function of  other clients' strategies $\boldsymbol{x}_{-n}$.

\subsection{Equilibrium Existence}\label{equilibrium_existence}
We first aim to understand whether an NE of Game \ref{game-formulation} exists. 
We note that,  without any assumption of the accuracy function $A(\cdot)$, the profit function $\Pi(\cdot)$, the mechanism $g_n(\cdot)$, and the cost $C_n^{\rm pri}(\cdot)$, the game analysis is intractable. There may not exist an NE, and in general, characterizing NE is NP-hard \cite{roughgarden2010algorithmic}. However, under some mild assumptions, we can guarantee the existence of an NE.
\begin{assumption}\label{concavity}
$A(\boldsymbol{s}, \boldsymbol{\epsilon})$ concavely increases in $s_n$, $\forall n$. 
%and $C_n^{\rm pri}(s_n)$ is a convex increasing function in $s_n$, $\forall n$.
\end{assumption}
Assumption \ref{concavity} is a widely adopted assumption (e.g., \cite{zhan2020learning,karimireddy2022mechanisms}), meaning that more data contribution leads to a smaller marginal accuracy improvement. Also, our experiments using the CIFAR-10 dataset show that the global model accuracy exhibits a concave increasing trend in the data contribution level (see Fig.~\ref{accuracy-data} in Sec. \ref{validation}).
\begin{assumption}\label{concaveity-2}
%	$g''_n\Pi+2g_n'\Pi'+g_n\Pi''\le 0$, $\forall n$.
$\frac{\partial^2 g_n}{\partial {s_n}^2}\cdot \Pi + 2\frac{\partial g_n}{\partial s_n} \cdot \frac{\partial \Pi}{\partial A}+g_n \cdot  \frac{\partial^2 \Pi}{\partial A^2}\le 0$, $\forall n$.
\end{assumption}
Assumption \ref{concaveity-2} is a sufficient condition to ensure that a client's allocated profit concavely increases in its data contribution level. This assumption holds for various practical profit functions (e.g., linear ones) and allocation mechanisms (e.g., egalitarian and sublinear ones) \cite{yu2020fairness}.
%\footnote{Please refer to the supplementary material where we use various concrete function forms to justify the assumption.}
\begin{assumption}\label{convexity}
$C_n^{\rm pri}(s_n)$ convexly increases in $s_n$, $\forall n$. 
\end{assumption}
Assumption \ref{convexity} implies that as a client contributes more data in FL training, it will experience more significant increments in privacy cost. It can, for example, model the linear cost function \cite{karimireddy2022mechanisms,hu2020trading}, which characterizes the risk associated with the privacy leakage in revealing $s_n$ data points.

With the above assumptions, we show that an NE exists.
\begin{theorem}\label{existence}
Under Assumptions \ref{concavity}-\ref{convexity}, there exists a Nash equilibrium of Game \ref{game-formulation}.
\end{theorem}
\textit{Due to space limitations, we defer all the technical proofs to the supplementary material.}

%Theorem \ref{existence} is based on the conclusion in \cite{osborne1994course} where there exists a NE if the game has a finite set of players, and the players' payoffs are quasi-concave in their strategies. 
Theorem \ref{existence} provides an important result where there exists a stable outcome of clients' interactions. In practice, clients are expected to play the NE strategies as no client can be better off by unilaterally deviating from the equilibrium.  
%Note that this is an important result, as clients are likely to play stable equilibrium strategies in practice.

\subsection{Equilibrium Properties}\label{equilibrium_properties}
Having established the NE existence in Theorem \ref{existence}, we now try to characterize the NE. However, it is still difficult to give a closed-form equilibrium characterization mainly due to the lack of specific function forms for $A(\cdot)$, $\Pi(\cdot)$, $g(\cdot)$, and $C_n^{\rm pri}(\cdot)$. Nevertheless, we are able to analyze the monotonicity property of the NE.
\begin{theorem}\label{Impact-mu-D}
Under Assumptions \ref{concavity}-\ref{convexity},  a client's equilibrium $s_n^*$ is non-increasing in $\mu_n$ and non-decreasing in $D_n$, $\forall n$.
\end{theorem}
\textbf{Insights}: Theorem \ref{Impact-mu-D} has two implications. First, if a client is more privacy sensitive, it will use less data for FL training to reduce privacy costs. Second, having more data corresponds to greater flexibility for decision-making, and a client can choose to contribute more to earn more profits.

Besides Theorem \ref{Impact-mu-D}, we are particularly interested in how data quality (i.e., $\boldsymbol{\epsilon}$) affects the clients' equilibrium. To this end, we need to make another mild assumption. 
\begin{assumption}\label{assumption}
$\forall h \in \{A, \{g_n\}_{n \in \mathcal{N}}\}\}$, we have 
\begin{equation}\label{inequalities}
\frac{\partial h}{\partial \epsilon_n}\le 0, \quad \text{and} \quad \frac{\partial^2 h}{\partial s_n \partial \epsilon_n}\le 0.
\end{equation}
\end{assumption} 
The first inequality in (\ref{inequalities}) means as a client has a lower data quality (i.e., a larger $\epsilon_n$), the global model accuracy $A$ and the profit share $g_n$ will decrease. The second inequality means that both the accuracy and profit share have a smaller  improvement in $s_n$ as a client's data quality decreases. Our experiments using CIFAR-10 (e.g., Fig.~\ref{accuracy-quality} in Sec. \ref{numerical}) are consistent with the assumption.
%\footnote{Please refer to the supplementary material where we use various functions of $A(\cdot)$ and $g_n(\cdot)$ as further justifications for (\ref{inequalities}).} 

Now, we are ready to characterize the impact of data quality on the clients' equilibrium behaviors.
\begin{theorem}\label{impact-quality}
Under Assumptions \ref{concavity}-\ref{assumption}, each client's equilibrium $s_n^*$ is non-increasing in $\epsilon_n$.
\end{theorem}
\textbf{Insights}: Theorem \ref{impact-quality} implies that as a client has fewer noisy labels (i.e., a smaller $\epsilon_n$), it will use more data for FL. Contributing  data of higher quality leads to a better global model and a larger profit share. This incentivizes more high-quality data contributions from clients. 

We would like to emphasize the significance  and generality of our theoretical results in Theorems \ref{existence}-\ref{impact-quality}:
\begin{itemize}
\item Our results do not rely on any particular function forms of global model accuracy, allocation mechanism, or privacy cost. Instead, our analysis is built on mild assumptions that apply to various function forms. Furthermore, our numerical results in Sec. \ref{numerical-NE} validate our theoretical results even when certain assumptions are not satisfied. 
\item Our results apply to highly heterogeneous clients, i.e., clients may vary in data qualities $\epsilon_n$, privacy sensitivities $\mu_n$, and data capacities $D_n$. This corresponds to practical scenarios in cross-silo FL applications.
%\item Our results are consistent with the experimental results in Sect 
\end{itemize}
\subsection{Equilibrium Calculation}\label{equilibrium_calculation}
We present a best response algorithm to calculate NE. 
%In this subsection, we give theoretical analysis to Game \ref{game-formulation}.
\begin{algorithm}[tb]
\caption{Best Response Update} 
\label{BRalgorithm}
\begin{algorithmic}[1]
\STATE {\bfseries Initialization} Let the iteration index be $t=0$. Each client $n\in \mathcal{N}$ starts with $s_n(t=0)=s_n^{\rm ini}$.
\REPEAT
\FOR{each client $n \in \mathcal{N}$}
\STATE {\bfseries Best response update:} client $n$ updates its data contribution level via solving:
\begin{displaymath}
s_n(t) = \min \left[ \argmax_{s_n \in \{1,2,\cdots, D_n\}} U_n(s_n, \boldsymbol{s}_{-n}(t))\right].
\end{displaymath}
\ENDFOR
\STATE Update strategy profile: $\boldsymbol{s}(t+1) \leftarrow\boldsymbol{s}(t)$.
\STATE Update iteration index: $t\leftarrow t+1$.
\UNTIL{$\boldsymbol{s}(t)$ converges.}
\end{algorithmic}
\end{algorithm}
The proposed algorithm is given in Algorithm \ref{BRalgorithm} 1, where the clients iteratively update their data contribution levels until the algorithm converges. Let $t\in \mathcal{Z}_{+}$ denote the iteration index. Each client $n$ starts with an initial decision $s_n^{\rm ini}$ (Line 1).
%\footnote{We do not consider the minimum data size to be zero due to two reasons. First, the clients have to use non-zero data points to train a local model. Second, a zero data size for all clients can lead to the singularity issue when computing the payoff functions.}
%Note that different from the formulation in (\ref{workerDecision}), we consider that the data size $x_n$ takes discrete values in $\{1,2,\cdots,D_n\}$. There are two advantages of using  discretization. First, we no longer face a non-convex problem with continuous data sizes. Instead, we are solving an integer program with finite decision space $\{1,2,\cdots, D_n\}$. Second,  discretizing is more reasonable in practice, as one cannot use a fraction of data points for model training.   
While the algorithm does not converge, each client $n$ updates its data contribution level $s_n(t)$ based on the best response dynamics (Line 4). 
Then, after all the clients update their strategies, the algorithm proceeds into the next iteration $t+1$. The algorithm terminates when the absolute difference between $s_n(t+1)$ and $s_n(t)$ for all $n\in \mathcal{N}$ is smaller than a predefined threshold $\tau$.

Note that under mild assumptions, we can prove that Algorithm 1 converges to a Nash equilibrium. Also, in all of our numerical experiments, we find that the algorithm converges quickly, e.g., within ten iterations. Please refer to our supplementary material for technical proof and numerical illustrations of the algorithm convergence.

%We use an example below to show that the inequalities in (\ref{inequalities}) hold.
%\begin{Exam}
%	Consider a model accuracy decomposition:
%	\begin{equation}
%		A(\boldsymbol{s}, \boldsymbol{\epsilon})=A(\boldsymbol{s},\boldsymbol{\epsilon})|_{\boldsymbol{\epsilon}=\boldsymbol{0}}- r
%	\end{equation}
%\end{Exam}

\section{Numerical Results}\label{numerical}
We conduct numerical experiments to validate our assumptions and analysis. Specifically, in Section \ref{experiment-setup}, we discuss the experimental setup. In Section \ref{validation}, we train FL models to validate our assumptions. In Section \ref{numerical-NE}, we study the equilibrium results (calculated by Algorithm 1) under various practical mechanisms and obtain more insights.  \textbf{We have included the codes in the supplementary material.}

\subsection{Experimental Setup}\label{experiment-setup}
Since the client number in cross-silo FL is usually small, we consider $5$ clients who participate in the entire FL process.  We use CIFAR-10 \cite{cifar10} to train FL models using FedAvg. The CIFAR-10 dataset contains $10$ classes with $5*10^4$ training data and $10^4$ test data. We uniformly randomly split the training data among the $5$ clients and hence $D_n=10^4, \forall n$. The key hyper-parameters are as follows. We use ResNet-18 \cite{resnet} as the model structure. We set the local epoch number as 5 and batch size as 64. We set the local and global learning rates to be $0.1$ and $1$, respectively.
%\footnote{Please refer to our supplementary material on the hyper-parameter tuning process.}

Next, we discuss how we generate data with different qualities. For each client $n$, with a probability $\epsilon_n$, we randomly reassign each data point with a different label (from the remaining $9$ classes). As a result, each client's reassigned dataset statistically has an $\epsilon_n$ proportion of wrong labels compared to the original CIFAR-10 dataset. 

%\begin{table}[t]
%	\caption{Key hyper-parameters.} \label{hyper-parameters}
%	\begin{center}
%		\begin{tabular}{c|c}
%			\textbf{Name}  &\textbf{Description} \\
%			\hline \\
%		%	Client number   & 5\\
%		%	Aggregation algorithm  & FedAvg\\
%			Model  structure     &ResNet-18 \\
%		    Local epoch number       &5 \\
%			Local batch size            &64\\
%		    Local learning rate         &0.1\\
%		    Global learning rate        & 1\\
%		   % Communication rounds        & 50\\
%		    \hline 
%		\end{tabular}
%	\end{center}
%\end{table}

\begin{figure}[t]
\vspace{-5mm}
\centering
\subfloat[Global model accuracy vs. communication rounds. ]{\includegraphics[width=1.59in]{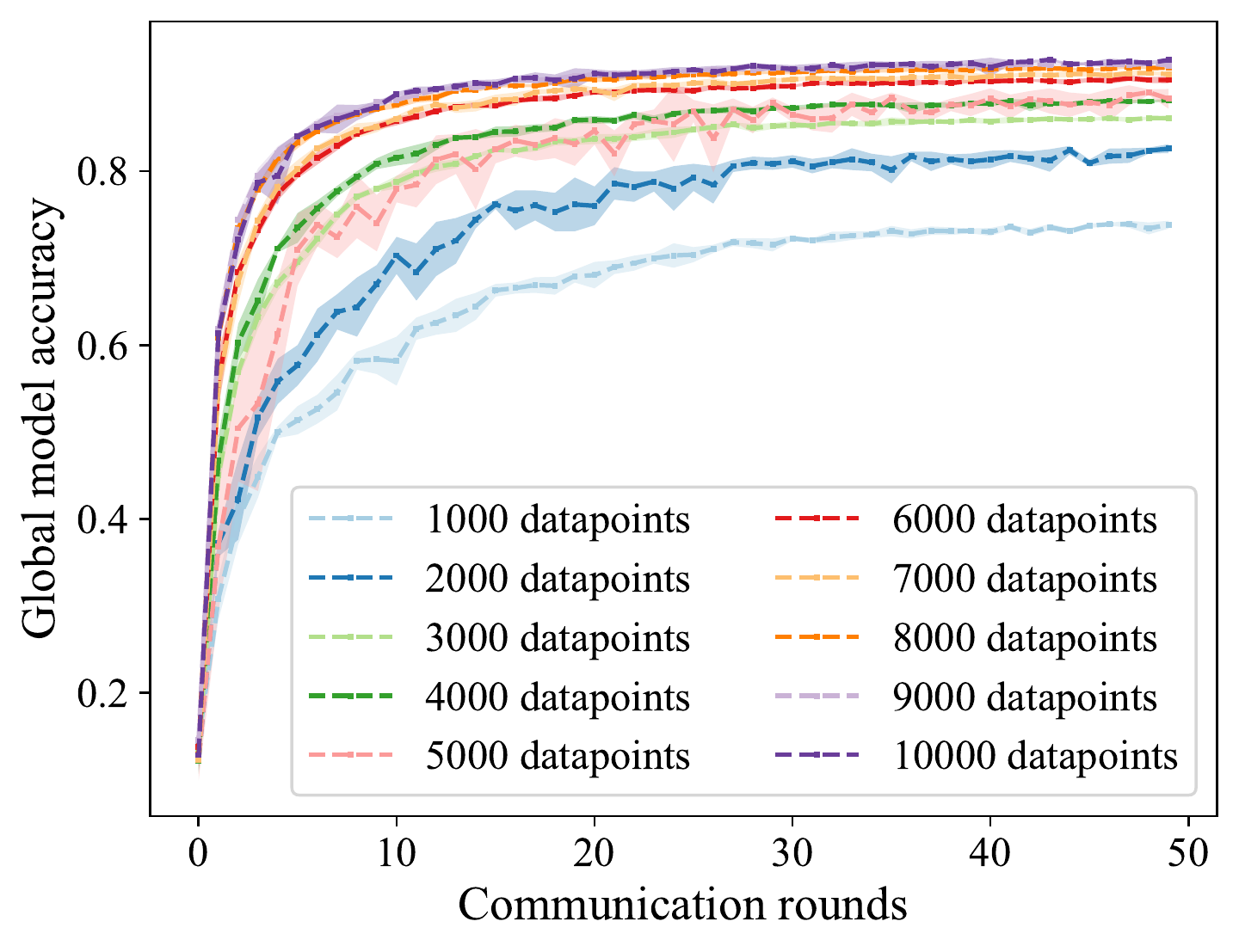}
\label{accuracy-rounds}}
\hfil
\subfloat[Global model accuracy vs. data contribution at round $50$. ]{\includegraphics[width=1.59in]{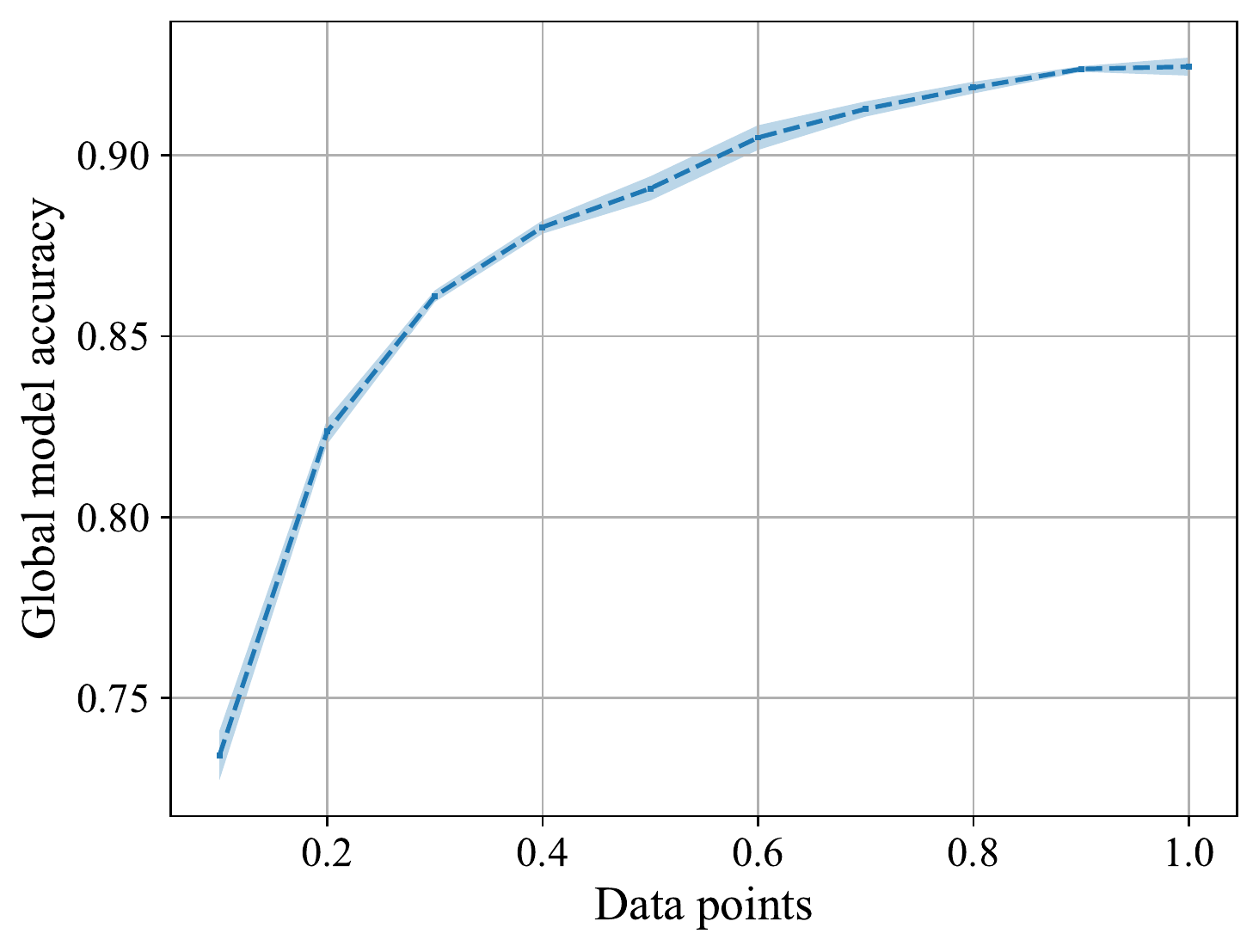}
\label{accuracy-data}}
\hfil
\caption{Impact of data contribution on global model accuracy.}
\label{FL-data-size}
\vspace{-3mm}
\end{figure}

\subsection{Impact of Data Contribution Level and Data Quality on Global Model Accuracy}\label{validation}
We conduct experiments to validate our technical assumptions. In particular, we study how the clients' data contribution $\boldsymbol{s}$ and data quality $\boldsymbol{\epsilon}$ affect the global model accuracy.

\textbf{Impact of data contribution on global model accuracy}: 
%We first investigate how clients' data contribution affects the global model accuracy (see Fig.~ \ref{FL-data-size}). 
Fig.~\ref{FL-data-size} plots how the global model accuracy depends on the clients' data contribution.
In the experiments, each client uses a randomly selected dataset with size $s_n\in 10^4\cdot \{0.1, 0.2, \cdots, 1\}$, and we consider clients have clean labels (i.e., $\epsilon_n=0$, $\forall n$). We repeat each experiment 5 runs. 
%Fig.~ \ref{accuracy-rounds} plots how the global model accuracy depends on the communication rounds under different data contribution levels. Fig.~ \ref{accuracy-data} plots how the global model accuracy changes with data contribution at round $50$.

In Fig.~\ref{accuracy-rounds}, we 
%observe that more data contribution leads to a faster global model convergence (since clients' data are clean).  Second, we 
observe that given a communication round (e.g., round $30$), the model accuracy increases in the data contribution levels. Further, we observe in Fig.~\ref{accuracy-data} that the global model accuracy exhibits a concavely-like increasing trend in the clients' data contribution, which is consistent with Assumption \ref{concavity}. In fact, the study in \cite{zhan2020learning} also revealed an accuracy curve that concavely increases in the clients' data contributions using the MNIST dataset. 

\begin{figure}[t]
\vspace{-5mm}
\centering
\subfloat[Global model accuracy vs. communication rounds. ]{\includegraphics[width=1.59in]{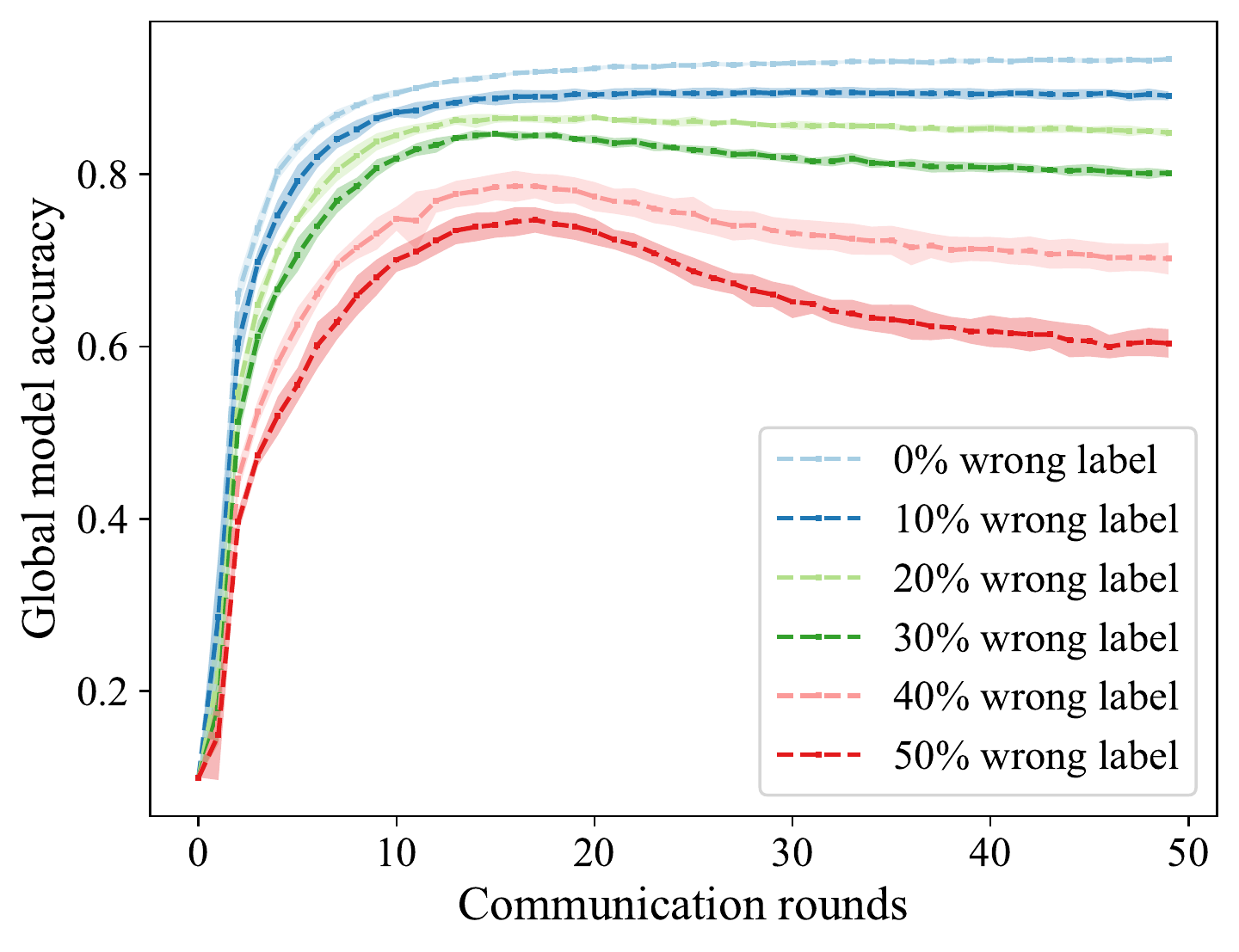}
\label{accuracy-rounds-2}}
\hfil
\subfloat[Global model accuracy vs. data quality at round $50$. ]{\includegraphics[width=1.59in]{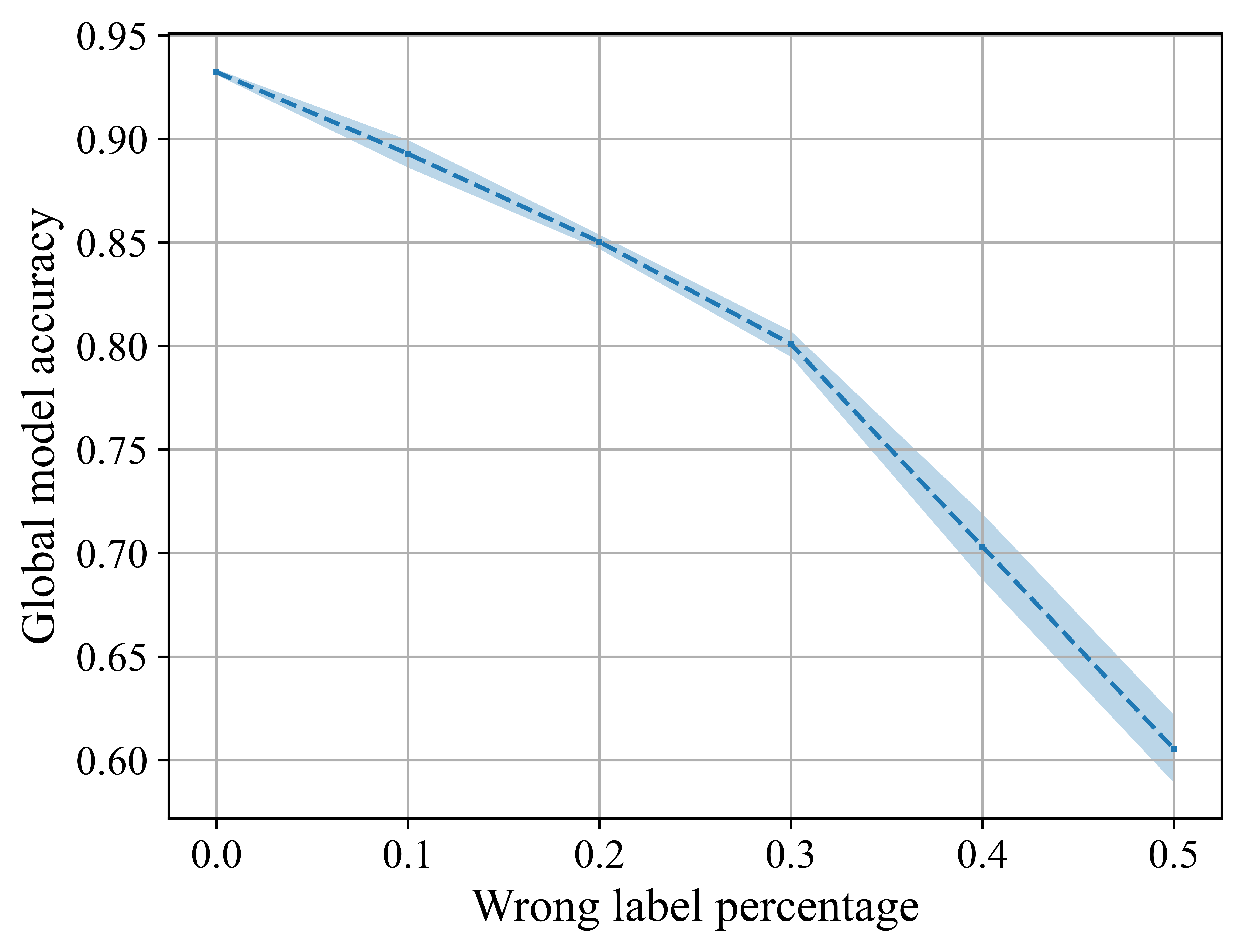}
\label{accuracy-quality}}
\hfil
\caption{Impact of data quality on global model accuracy.}
\label{FL-data-quality}
\vspace{-4mm}
\end{figure}

\textbf{Impact of data quality on global model accuracy}:
%Next, we investigate how the clients' data quality influences the global model accuracy (see Fig.~ \ref{FL-data-quality}).
Fig.~\ref{FL-data-quality} plots how the global model accuracy changes with clients' data quality. In the experiments, each client uses all local data for FL training, and the clients' data quality (i.e., wrong label percentage $\epsilon_n$) takes values in  $\{0, 0.1, 0.2, \cdots, 0.5\}$.
%\footnote{We conduct more experiments in the supplemental material with $\epsilon_n\in\{0.01, 0.02, 0.05\}$ and the results continue to hold.}
 We repeat each experiment 5 times. 
%Fig.~ \ref{accuracy-rounds-2} plots how the global model accuracy depends on the communication rounds under different $\epsilon_n$. Fig.~ \ref{accuracy-quality} plots how the global model accuracy changes with $\epsilon_n$ at round $15$.

In Fig.~\ref{accuracy-rounds-2}, we observe that when clients have relatively high-quality data (e.g., $\epsilon_n\in\{0,0.1\}$), the global model accuracy improves as the FL training proceeds. However, when clients' data quality degrades (e.g., $\epsilon_n\ge 0.2$), the global model tends to overfit with more communication rounds. For example, at $\epsilon_n=0.5$, the global model accuracy begins to decrease after round $15$. Deep neural networks (e.g., ResNet) tend to overfit to label noise, and overfitting can escalate in FL where clients' data are private and distributed \cite{fang2022robust}. This interesting observation bears important implications on how label noise affects FL, and it calls for FL algorithm improvement to mitigate potential overfitting due to label noise.

In Fig.~\ref{accuracy-quality}, we observe that the global model accuracy decreases in $\epsilon_n$.  As clients' data have more noisy labels, the learned global model would have  worse performance. Also, an empirical study in \cite{xu2022fedcorr} found that label noise can degrade the global model performance. These observations are consistent with Assumption \ref{assumption}. 

We summarize the key observations in Figs. \ref{FL-data-size}-\ref{FL-data-quality} below:

\begin{observation}
(i) The global model accuracy concavely increases in the clients' data contributions, but it decreases in the wrong label percentage.
%	(ii) The global model accuracy decreases in the wrong label percentage.
(ii) The global model tends to overfit when the wrong label percentage is large.
\end{observation}

\subsection{Nash Equilibrium Results}\label{numerical-NE}
%In Subsection \ref{BestResponse}, we introduce a best response algorithm for computing the Nash equilibrium. 
%Next, we study the Nash equilibrium solutions (calculated by Algorithm 1). 
%In Subsection \ref{mechanisms_metrics}, we introduce four practical profit allocation mechanisms to be implemented. In Subsection \ref{NE-solutions}, 
We study how the Nash equilibrium results (calculated by Algorithm 1) depend on the clients' data quality and privacy sensitivity, followed by the comparison of the four mechanisms (see Eq. (\ref{eg})-(\ref{sv})). 

\textbf{Experimental setup}: 
%We investigate how data quality and privacy sensitivity  affect the clients' equilibrium data contribution. 
In the experiments, we use a linear privacy cost, i.e., $C_n^{\rm pri}(s_n)=\mu_n s_n$ \cite{karimireddy2022mechanisms}.   Note that it is empirically infeasible to construct a tabular correspondence between clients' data contribution $\boldsymbol{s}$ and the global model accuracy $A(\boldsymbol{s}, \boldsymbol{\epsilon})$, as this requires FL training for $(D_n)^N=(10^4)^{5}=10^{20}$ times. Instead, we use a surrogate accuracy function, i.e., $A(\boldsymbol{s}, \boldsymbol{\epsilon}) = A(\boldsymbol{s},\boldsymbol{0})-\gamma\sum_{n\in\mathcal{N}}\frac{\epsilon_n s_n}{\sum_{n'=1}^{N}s_{n'}}$, where $A(\boldsymbol{s},\boldsymbol{0})$ represents the model accuracy when clients have clean labels, and the term $\gamma\sum_{n\in\mathcal{N}}\frac{\epsilon_n s_n}{\sum_{n'=1}^{N}s_{n'}}$ characterizes the accuracy loss caused by label noise.  The parameter $\gamma$ and the function $A(\boldsymbol{s},\boldsymbol{0})$ are obtained using curve fitting, 
%in Fig.~\ref{accuracy-data} and %Fig.~\ref{accuracy-quality},
and in particular,  $A(\boldsymbol{s},\boldsymbol{0})= \alpha_1\log(\alpha_2\sum_{n\in \mathcal{N}}s_n+\alpha_3)+\alpha_4 \sum_{n\in \mathcal{N}}s_n+\alpha_5$.
Moreover, we use a convex profit function, i.e., $\Pi(A(\boldsymbol{s}, \boldsymbol{\epsilon}))=\beta_1 A^2(\boldsymbol{s}, \boldsymbol{\epsilon})+\beta_2$. It can model the scenario where the total profit increases faster as the global model accuracy improves. For example, one would expect that the profit earned via improving $A$ from $0.9$ to $0.95$ is greater than improving $A$ from $0.7$ to $0.75$.
%\footnote{Please refer to our supplementary material for the detailed curve fitting process. We also conduct more experiments using different functions of $C_n^{\rm pri}(\cdot)$ and $\Pi(\cdot)$.} We initialize clients' decisions to be $s_n^{\rm ini}=0.1\cdot D_n, \forall n$.
%In this subsection, we study how various system parameters and profit allocation mechanisms affect the Nash equilibrium solutions. 
%We use three metrics to evaluate the above mechanisms.
%\begin{itemize}
%	%\item \textit{Contributed data quality}: the concept is defined as 
%	%$\sum_{n\in \mathcal{N}} (1-\epsilon_n)s_n^*$, which reflects both the equilibrium data contribution level $s_n^*$ and clients' data quality $\epsilon_n$.  
%	\item \textit{Global model accuracy}: it reflects both the equilibrium data contribution level $\boldsymbol{s}^*$ and clients' data quality $\boldsymbol{\epsilon}$.
%	\item \textit{Social welfare}
%\end{itemize}
%1) the clients' data contribution levels; 2)  the global model accuracy; and 3) social welfare, defined as the summation of all clients' payoffs.
%\begin{itemize}
%	\item \textit{Clients' data contribution levels}:
%	\item \textit{Global model accuracy}: 
%	\item \textit{Social welfare}: it is defined as the summation of all clients' payoffs at equilibrium. 
%\end{itemize}

\begin{figure}[t]
\vspace{-5mm}
\centering
\subfloat[$\epsilon_1=\epsilon_2=\epsilon_3=0$. ]{\includegraphics[width=1.59in]{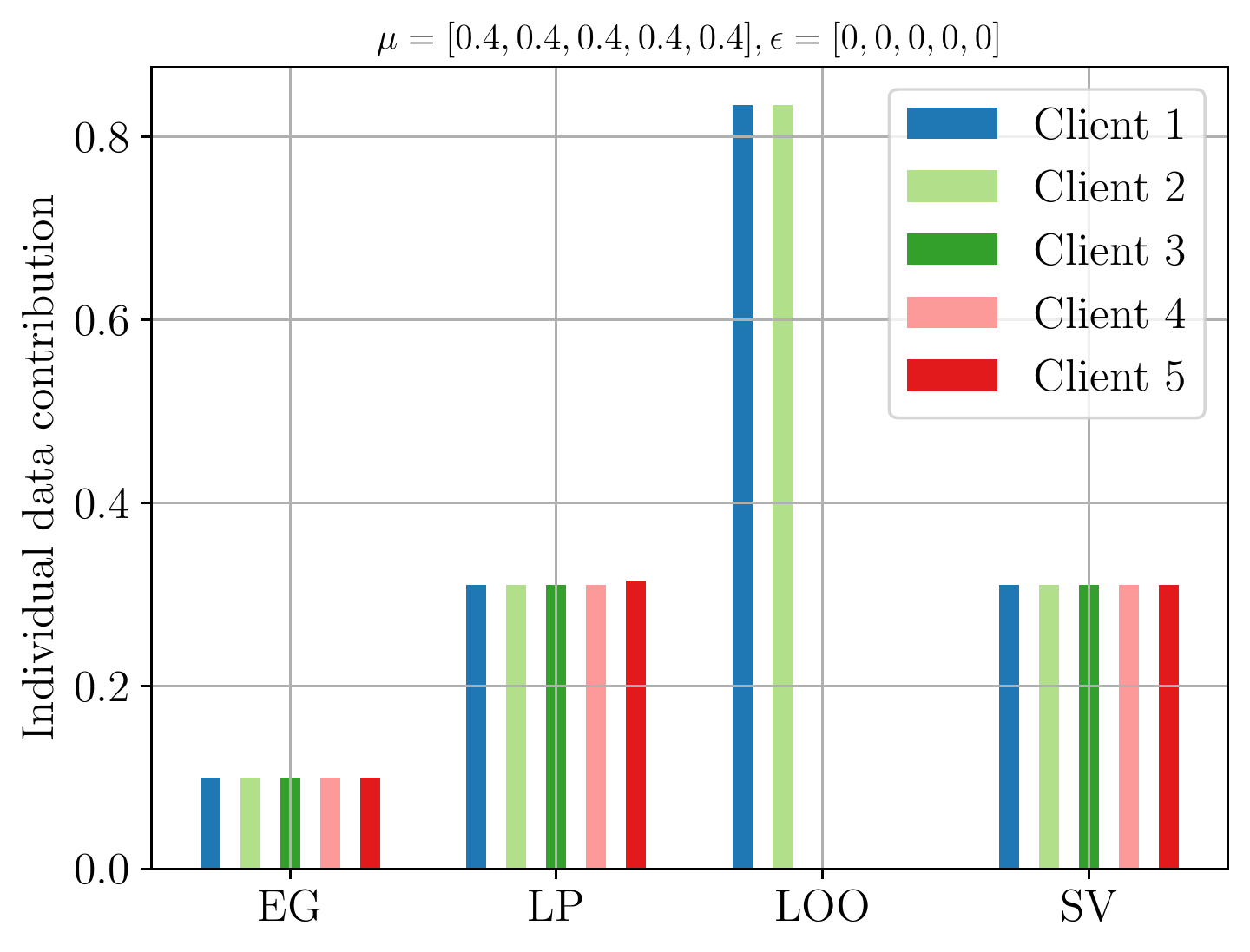}
\label{epsilon_0}}
\hfil
\subfloat[$\epsilon_1=\epsilon_2=\epsilon_3=0.1$. ]{\includegraphics[width=1.59in]{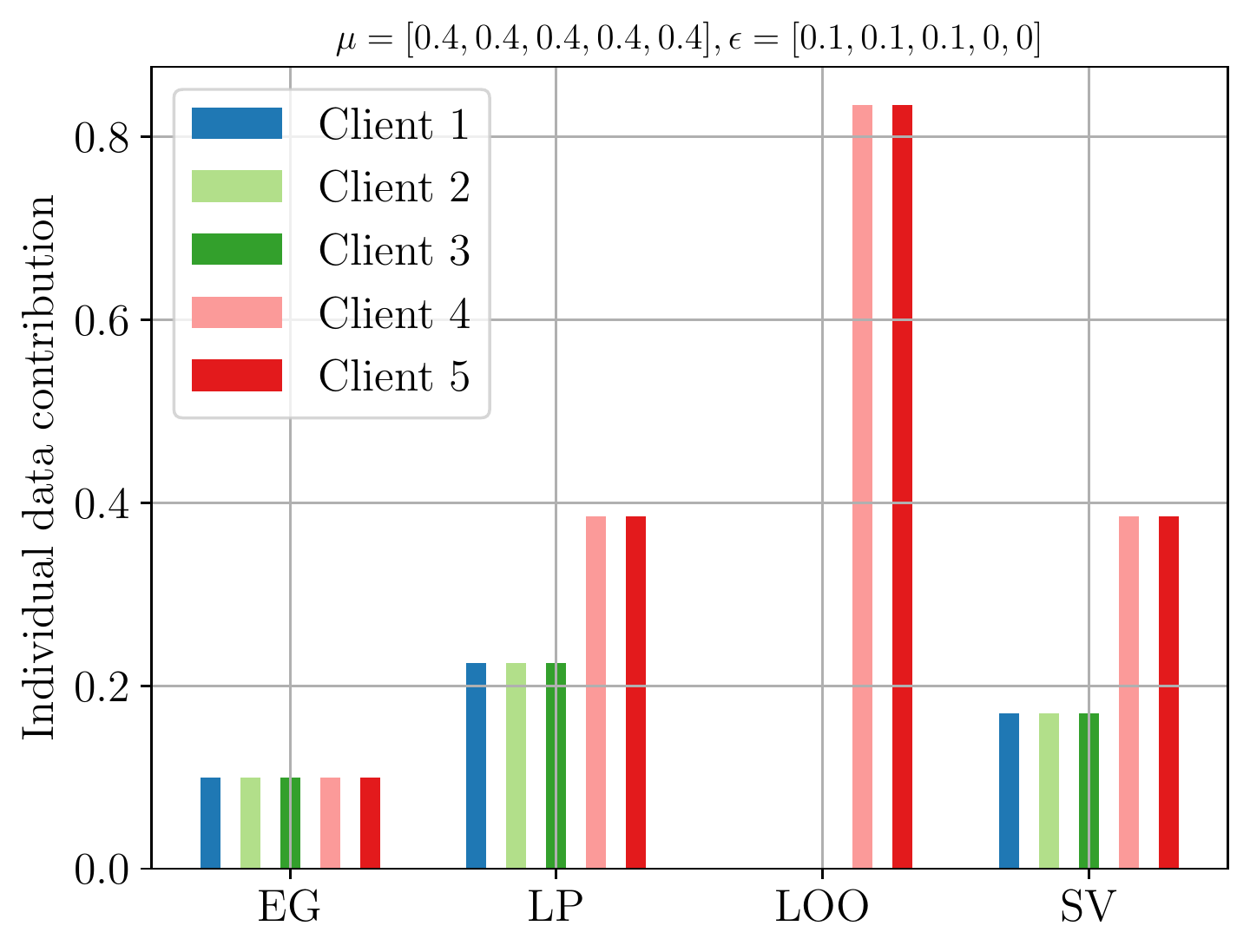}
\label{epsilon_01}}
\hfil
\subfloat[$\epsilon_1=\epsilon_2=\epsilon_3=0.3$. ]{\includegraphics[width=1.59in]{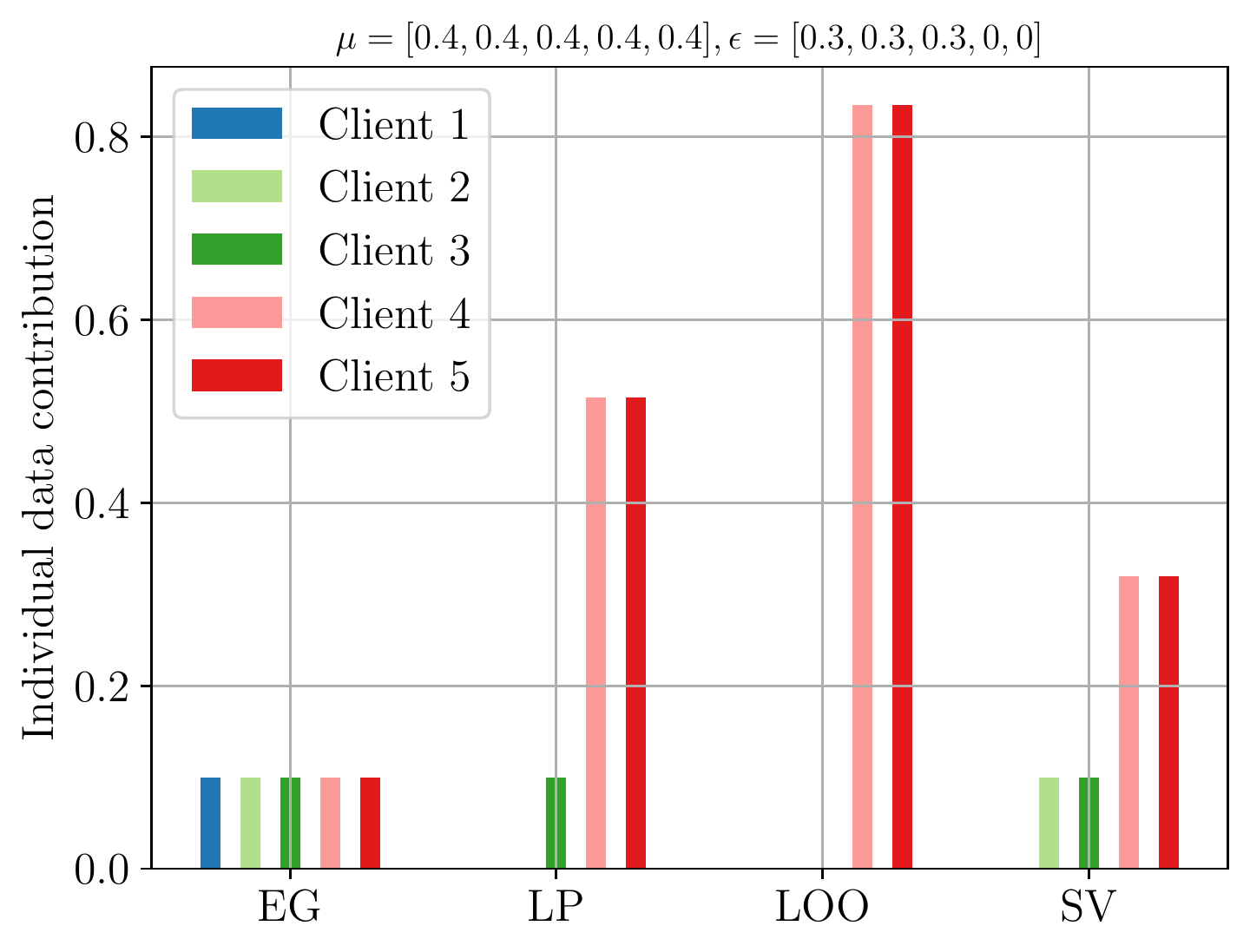}
\label{epsilon_03}}
\hfil
\subfloat[$\epsilon_1=\epsilon_2=\epsilon_3=0.4$. ]{\includegraphics[width=1.59in]{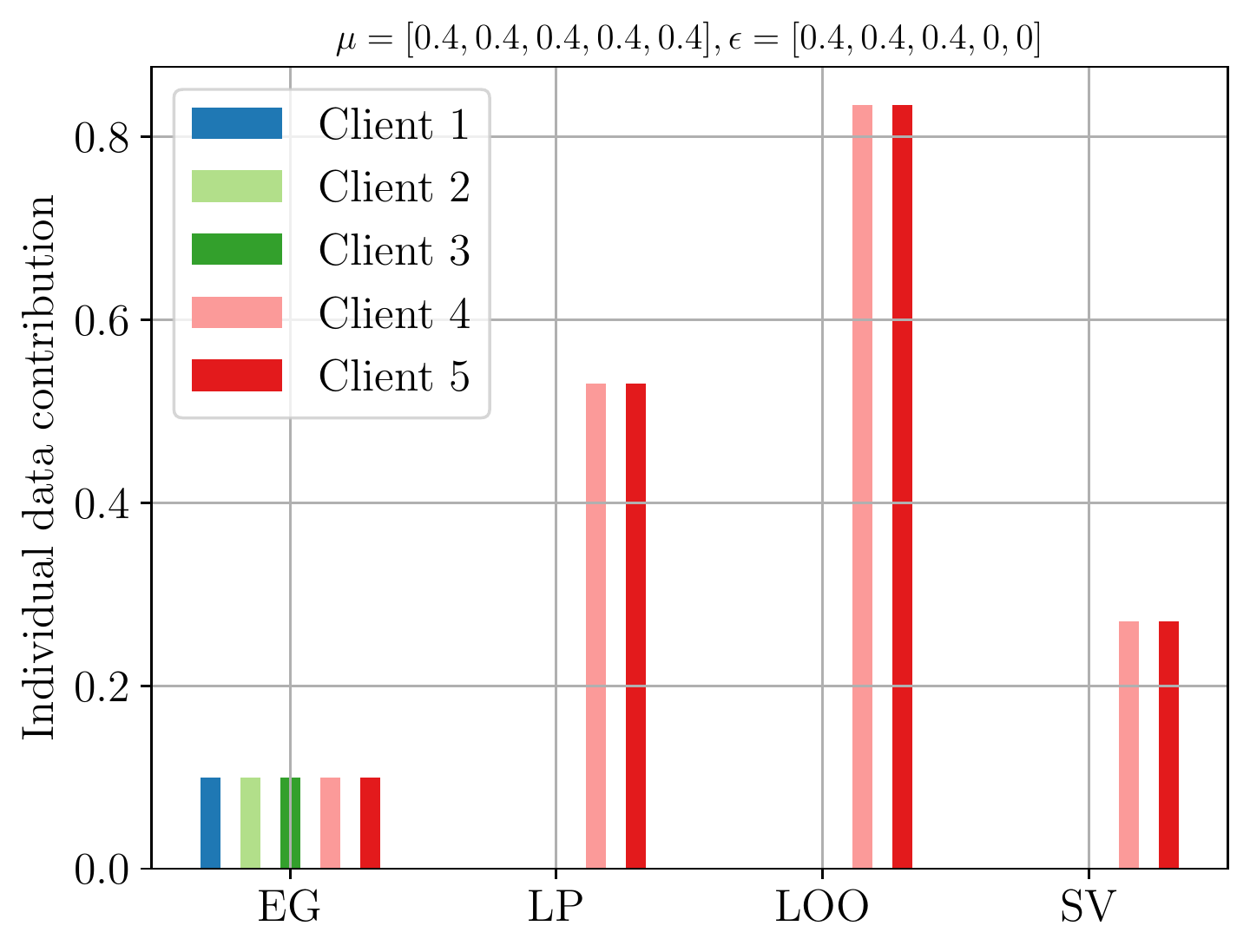}
\label{epsilon_04}}
\hfil
\caption{Impact of data quality on clients' individual data contributions at Nash equilibrium.}
\label{quality_individual}
\vspace{-3mm}
\end{figure}

\textbf{Impact of data quality}: We first investigate how data quality affects the clients' equilibrium data contributions $s_n^*$, $\forall n$. To this end, we consider that the clients have the same privacy sensitivity, i.e., $\mu_n=0.4, \forall n$. Clients have different data qualities, and in the experiments, we set $\epsilon_4=\epsilon_5=0$. We consider that clients 1-3 have the same wrong label percentage and it takes values in the set $\epsilon_1=\epsilon_2=\epsilon_3\in\{0, 0.1, 0.2, 0.3, 0.4, 0.5\}$. Fig.~\ref{quality_individual} plots how clients' equilibrium $s_n^*$ depends on $\epsilon_n$.
%\footnote{The equilibrium results for $\epsilon_n\in\{0.2, 0.5\}$ are very similar to cases when $\epsilon_n\in \{0.1, 0.4\}$, $\forall n\in\{1,2,3\}$. We defer these two cases to our supplementary material due to space limitations. }

In Fig.~\ref{quality_individual}, we observe that the data contribution of clients 1-3 (weakly) decreases in the wrong label percentage $\epsilon_n, \forall n\in \{1,2,3\}$, under all mechanisms. 
%client $n$'s equilibrium data contribution (weakly) decreases in its wrong label percentage $\epsilon_n (\forall n\in \{1,2,3\}$), under all profit mechanisms. 
This observation is consistent with Theorem \ref{impact-quality}.
Moreover, in Figs. \ref{epsilon_01}-\ref{epsilon_04}, we find that clients 4-5 use a no smaller data contribution than clients 1-3. This implies that under the above mechanisms, clients with higher-quality data tend to contribute more to FL. 
%As a client has more noisy labels (i.e., data of lower quality), the global model has a worse performance, leading to a lower profit share. This decentivizes low-quality data contribution and incentivizes high-quality data contribution. 
We summarize the observation below:

\begin{observation}
(i) A client's equilibrium data contribution decreases in its wrong label percentage. (ii) Clients with higher-quality data contribute more to FL at equilibrium.
\end{observation}

%\begin{figure}[t]
%	\centering
% \includegraphics[width=3in]{./figure/global_model_ne}
%		\label{global_model_ne}
%	\caption{Impact of data contribution on global model accuracy.}
%	%\vspace{-4mm}
%\end{figure}

\begin{figure}[t]
\vspace{-5mm}
\centering
\subfloat[$\mu_1=\mu_2=\mu_3=0.1$. ]{\includegraphics[width=1.59in]{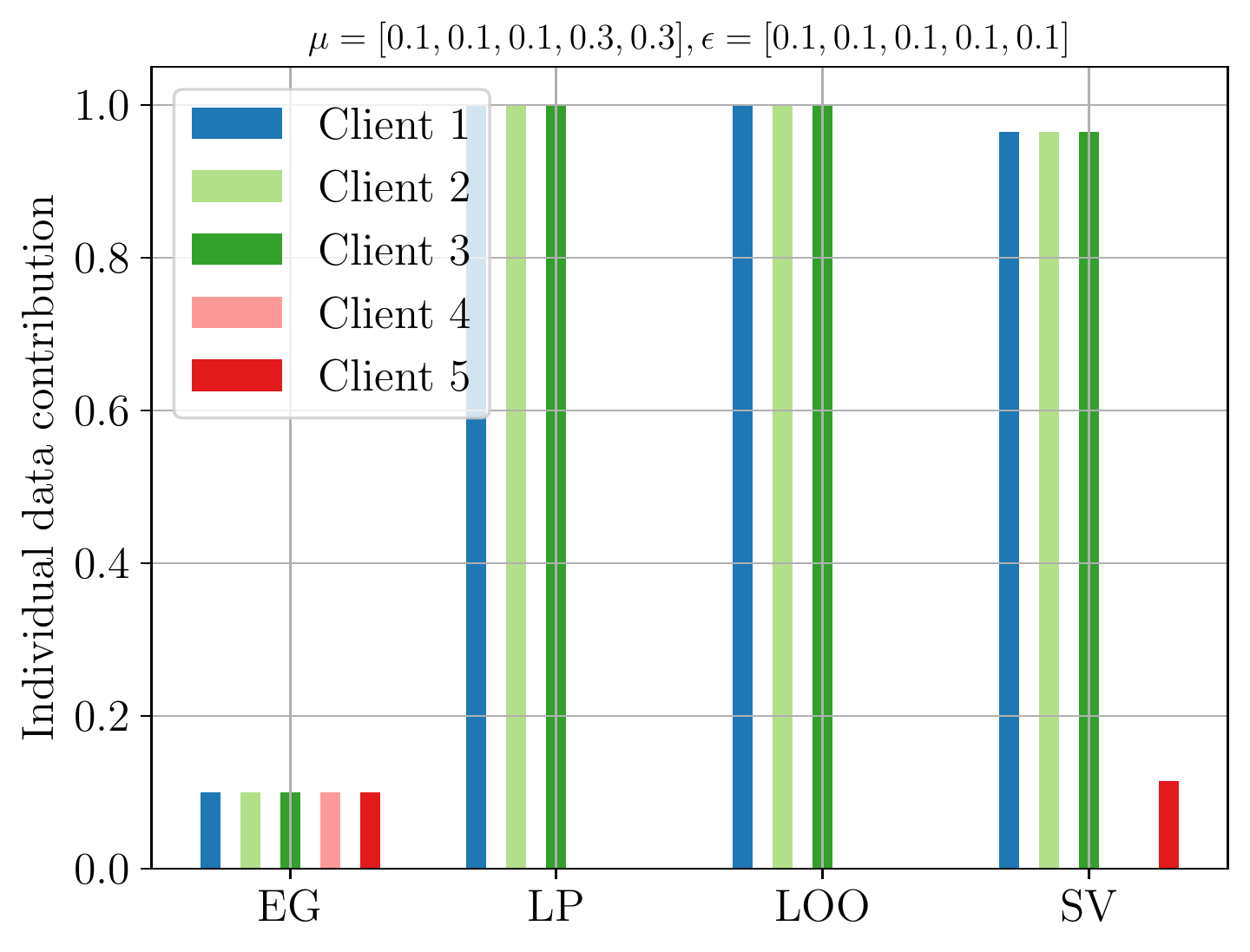}
\label{mu_01}}
\hfil
\subfloat[$\mu_1=\mu_2=\mu_3=0.3$. ]{\includegraphics[width=1.59in]{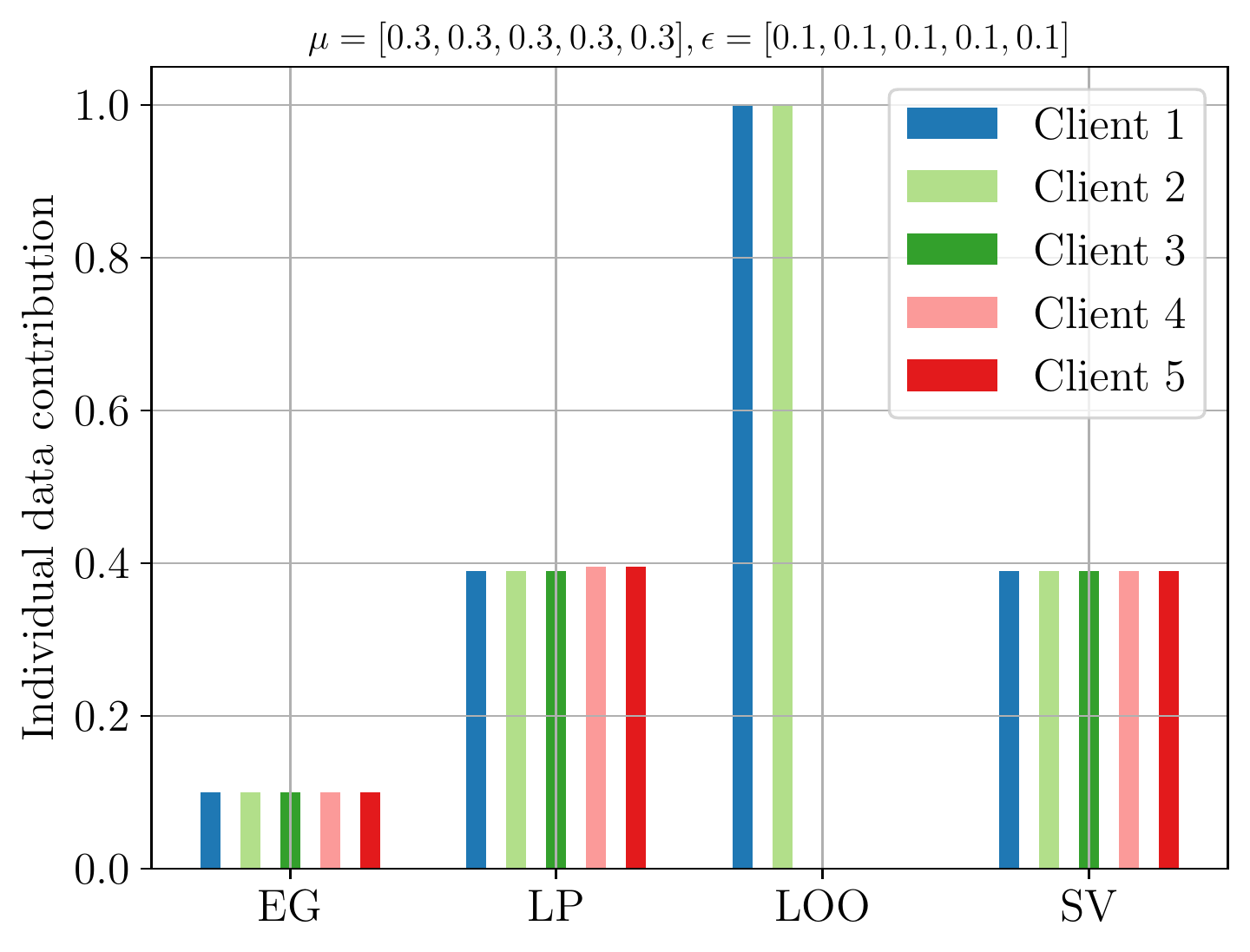}
\label{mu_03}}
\hfil
\subfloat[$\mu_1=\mu_2=\mu_3=0.5$. ]{\includegraphics[width=1.59in]{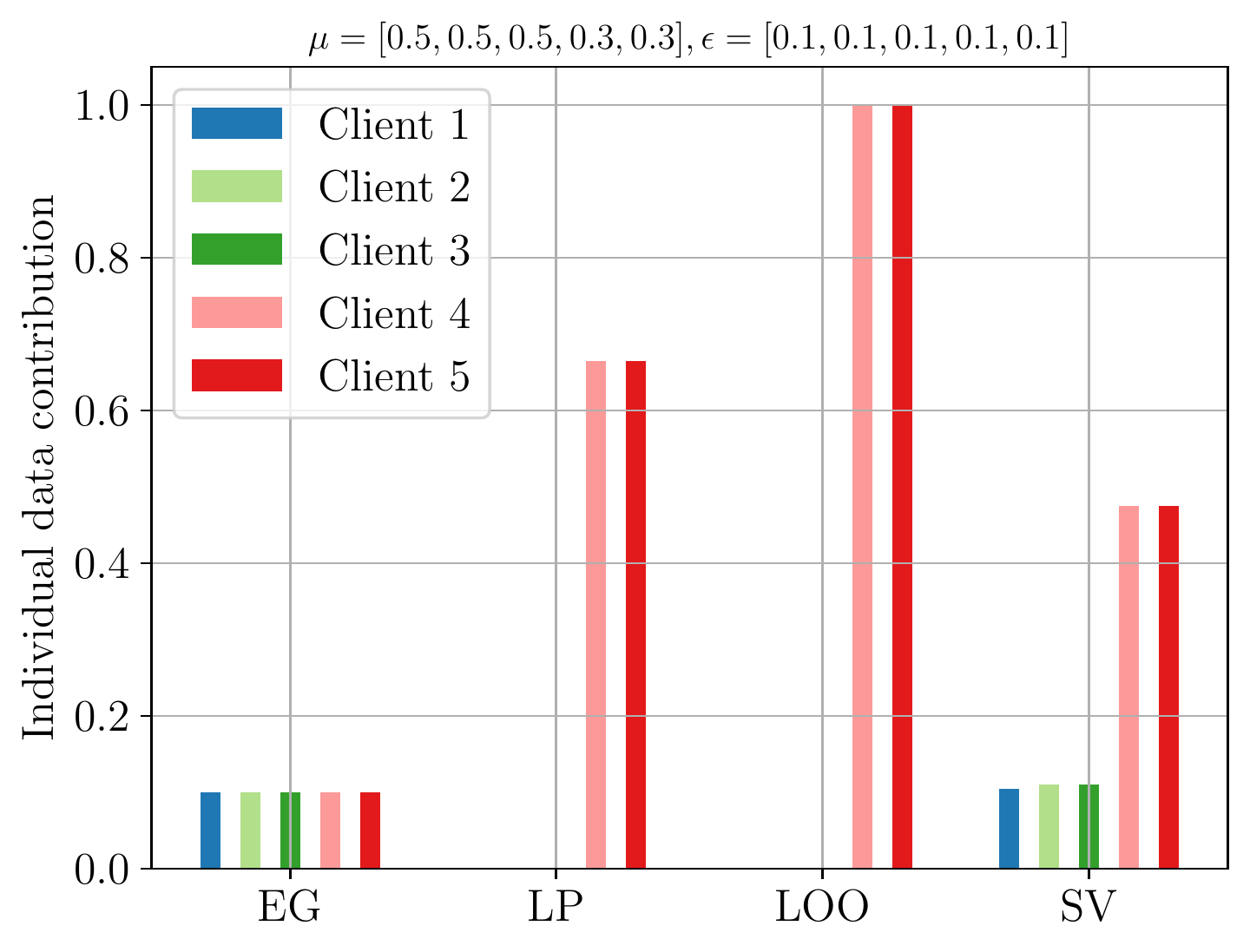}
\label{mu_05}}
\hfil
\subfloat[$\mu_1=\mu_2=\mu_3=0.9$. ]{\includegraphics[width=1.59in]{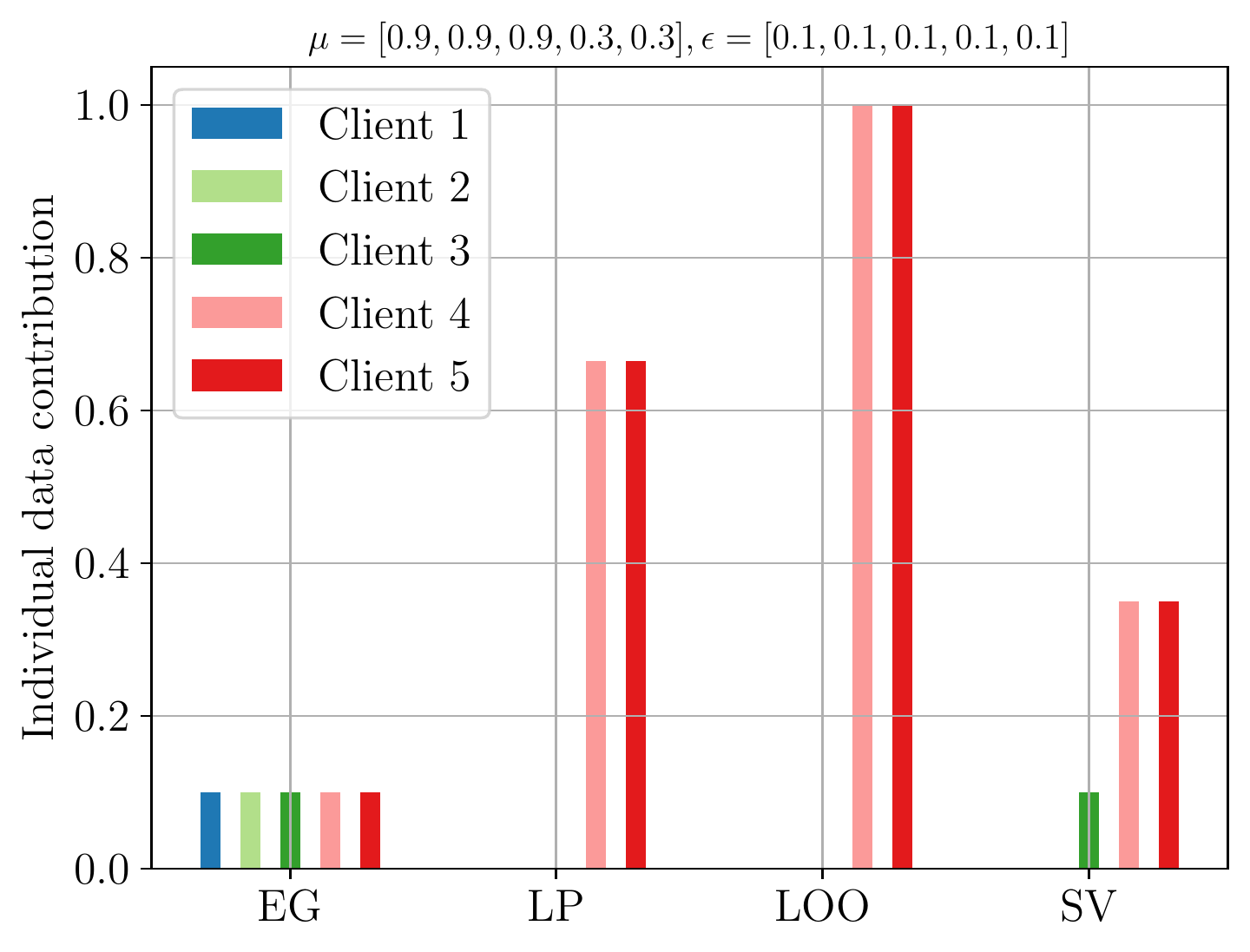}
\label{mu_07}}
\hfil
\caption{Impact of privacy sensitivity on clients' individual data contributions at Nash equilibrium.}
\label{privacy_individual}
\vspace{-3mm}
\end{figure}
%\footnote{For the ease of presentation}

\textbf{Impact of privacy sensitivity}: Next, we investigate how clients' privacy sensitivity affects the equilibrium data contributions. To this end, we consider that clients have the same data quality $\epsilon_n=0.1, \forall n$. Clients have different privacy sensitivities, and we set $\mu_4=\mu_5=0.3$. We consider that clients 1-3 have the same privacy sensitivity and it takes values  $\mu_1=\mu_2=\mu_3\in \{0.1, 0.3, 0.5, 0.7, 0.9\}$. Fig.~\ref{privacy_individual} plots how clients' equilibrium $s_n^*$ changes with $\mu_n$. 

In Fig.~\ref{privacy_individual}, we see that client $n$'s equilibrium data contribution weakly decreases in its privacy sensitivity $\mu_n, \forall n\in \{1,2,3\}$. This is consistent with Theorem \ref{Impact-mu-D}. Also, we observe that in Figs. \ref{mu_05} and \ref{mu_07}  that clients 4-5 contribute more data than clients 1-3, implying that clients with smaller privacy sensitivity tend to contribute more data at equilibrium. 
%As a client becomes more privacy sensitive, it incurs a larger privacy cost and is less likely to contribute more data for FL. 
We summarize the observations below:
\begin{observation}
(i)	A client's equilibrium data contribution decreases in its privacy sensitivity. (ii) Clients with smaller privacy sensitivity contribute more data at equilibrium.
\end{observation}

\textbf{Comparison of various mechanisms}: We now compare the four profit allocation mechanisms in terms of the clients' equilibrium data contribution. 

First, we observe that EG leads to the same equilibrium solutions for all the parameter settings in Fig.~\ref{quality_individual} and Fig.~\ref{privacy_individual}. EG cannot incentivize clients with high-quality data to contribute, as it always equally splits the total profit. Second, when clients have the same data quality (e.g., Fig.~\ref{epsilon_0}), we observe that LP and SV tend to induce similar data contributions among clients, while LOO leads to an imbalanced client contribution (i.e., clients 3-5 do not contribute). 
%LP and SV induce similar contributions since the clients have the same data quality (see Eq. (\ref{lp}) and Eq. (\ref{sv})). 
Under LOO, once some clients contribute enough data, they can harvest most of the profits and
decentivize other clients to contribute data, leading to imbalanced data contribution at equilibrium. Third, when clients differ in data quality (e.g., Figs. \ref{epsilon_01},\ref{epsilon_03},\ref{epsilon_04}), we find that LP, LOO, and SV can incentivize clients with higher-quality data to contribute more data than those with lower-quality data. This demonstrates the effectiveness of the three mechanisms in incentivizing high-quality data contribution in FL.  We summarize the above key observations as follows:

\begin{observation}
EG cannot incentivize high-quality data contribution, while LP, LOO, and SV incentivize more data contribution from clients with higher-quality data. 
\end{observation}

\begin{figure}[t]
\vspace{-5mm}
\centering
\subfloat[Global model accuracy vs. data quality. ]{\includegraphics[width=1.59in]{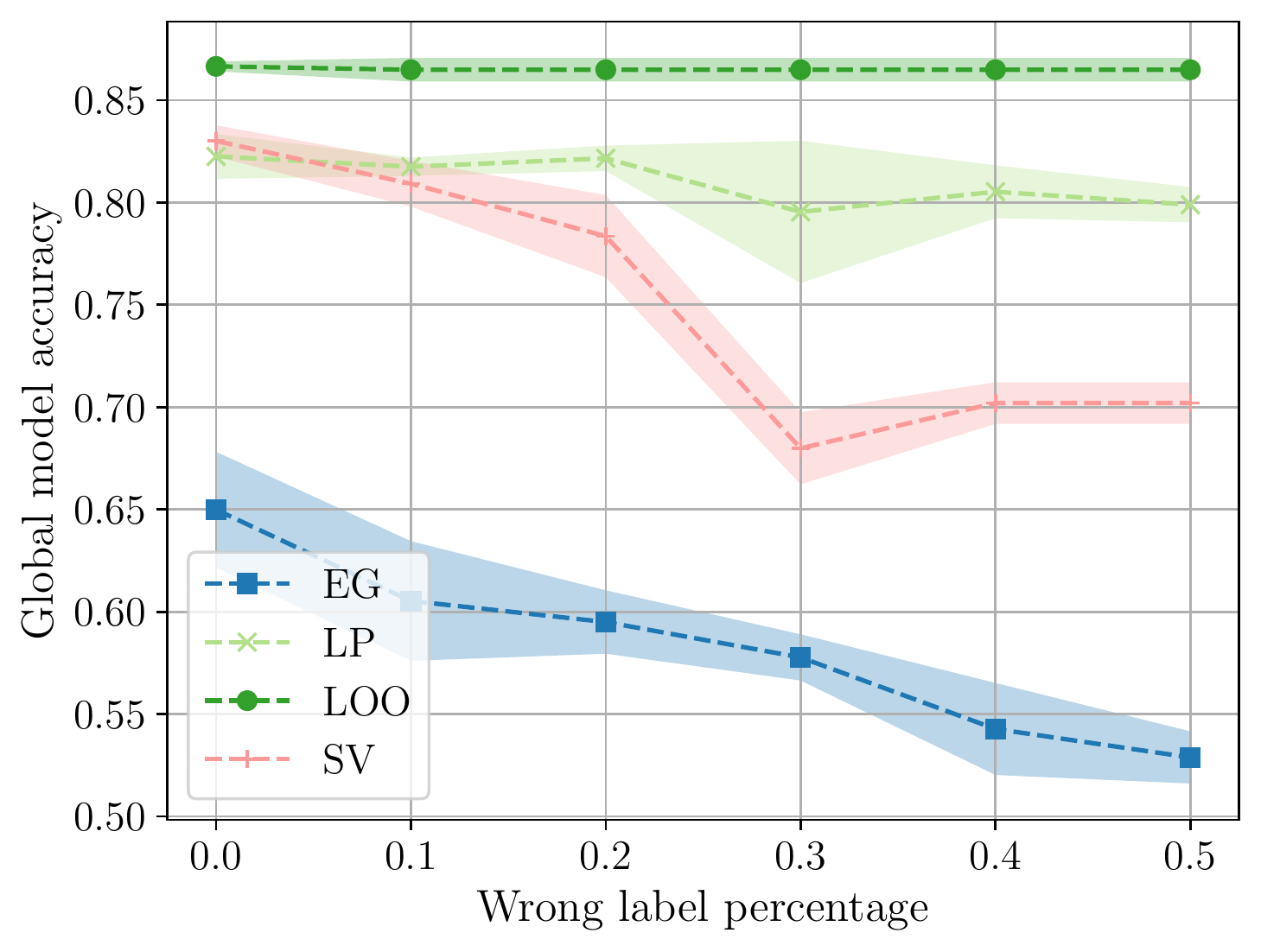}
\label{accuracy-ne-quality}}
\hfil
\subfloat[Global model accuracy vs. privacy sensitivity. ]{\includegraphics[width=1.59in]{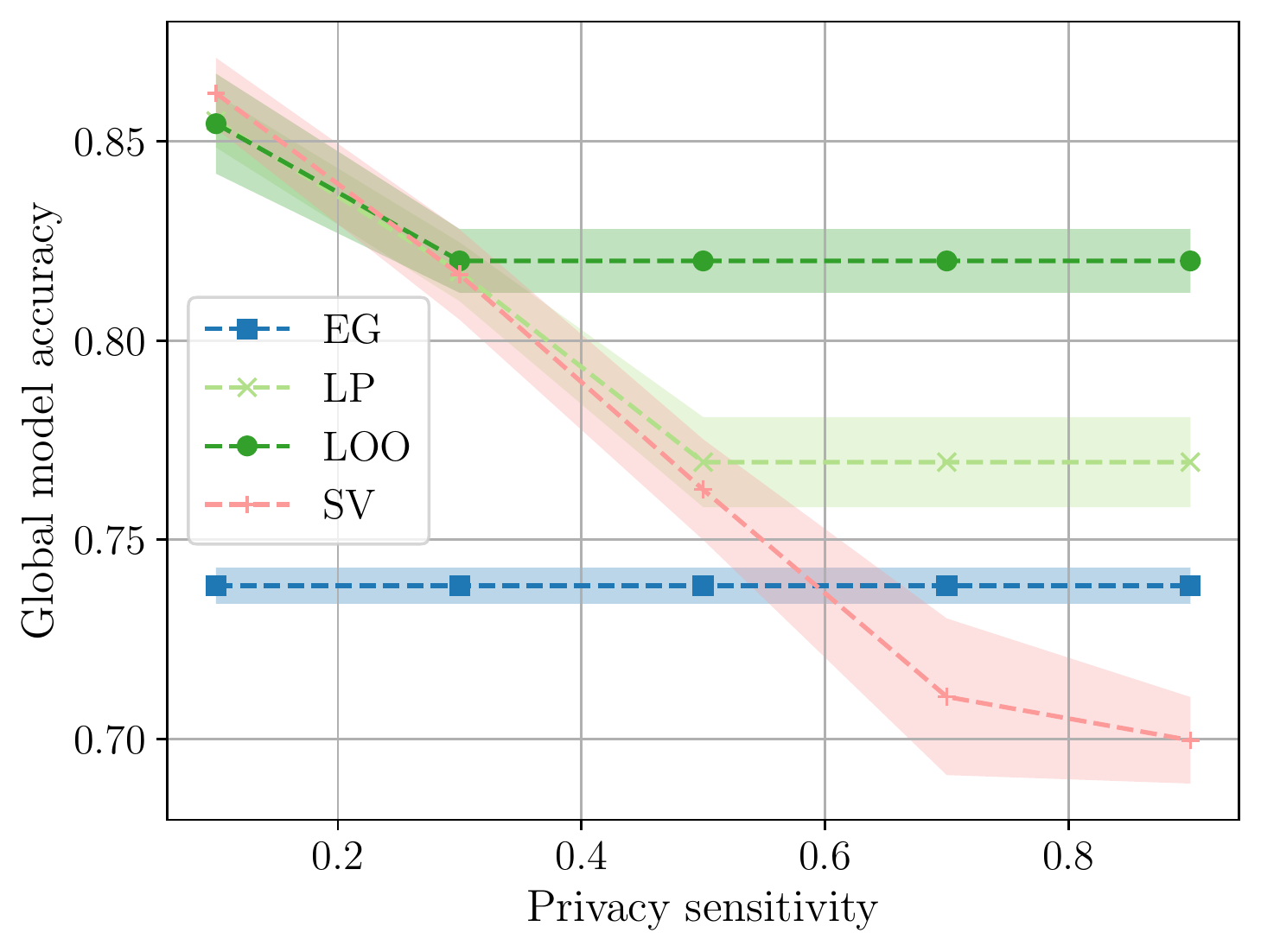}
\label{accuracy-ne-privacy}}
\hfil
\caption{Global model accuracy at equilibrium.}
\label{accuracy-ne}
\vspace{-5mm}
\end{figure}

To further compare the four mechanisms, we use the equilibrium results (e.g., in Fig.~\ref{quality_individual} and Fig.~\ref{privacy_individual}) to train FL models and present the global model accuracy. 
%use the hyper-parameters in Table \ref{hyper-parameters}, 
%repeat each experiment 5 times. 
Fig.~\ref{accuracy-ne} shows how the global model accuracy at equilibrium changes with data quality and privacy sensitivity.

In Fig. \ref{accuracy-ne-quality} and Fig.~\ref{accuracy-ne-privacy}, we observe that the global model accuracy generally decreases in the wrong label percentage and the privacy sensitivity. As clients have data of lower quality or become more privacy sensitive, they  tend to contribute less data, resulting in a worse global model. Furthermore, we find an interesting observation that LOO generally induces the best global model among the four mechanisms. As can be seen in Fig.~\ref{quality_individual} and Fig.~\ref{privacy_individual}, LOO tends to incentivize the most amount of total data contribution albeit from fewer clients with high-quality data (or small privacy sensitivity), leading to the best global model performance. We summarize this observation below.

\begin{observation}
LOO tends to outperform EG, LP, and SV in obtaining a better global model at equilibrium.
\end{observation}

%\section{Discussions}

%\subsection{Limitations and Possible Solutions}
%\subsection{Social Impact}

\vspace{-2mm}
\section{Discussions}\label{discussions}
Our paper proposes a first general framework to incentivize data contribution in cross-silo FL using profit allocation. However, some unaddressed challenges merit future study.

\textbf{Label imbalance}: Our work focuses on label noise in FL and how it affects clients' strategic data contribution. Another important problem in FL is label imbalance, which has drawn extensive recent research attention. However, its impact on clients' strategic data contribution is poorly understood. We plan to analyze the incentive design in cross-silo FL with label imbalance, and some preliminary numerical results are given in the supplementary material.

\textbf{Data duplication}: We made a widely adopted assumption that the server knows the size of clients' training dataset $s_n$ (i.e., data contribution). In practice, the server may not know this information, and clients can exaggerate their contributions by data duplication/cloning.  We conjecture that mechanisms such as LOO and SV can help detect and decentivize malicious data duplication. Under LOO and SV, training models with duplicated data is expected to make little contribution to the global model accuracy, leading to a small contribution index and profit share. This decentivizes data duplication from clients.

\textbf{Optimal mechanism design}: We evaluated our general framework using four widely adopted mechanisms, i.e., EG, LP, LOO, SV. This further motivates the optimal mechanism design that can potentially better incentivize data contribution. It is also interesting to incorporate fairness considerations which are critical in cross-silo settings with medical and financial applications.

%\textbf{Label imbalance in appendix}
\vspace{-3mm}
\section{Conclusion}\label{conclusions}
%In this paper, we study the incentive design in cross-silo FL with noisy labels. 
In cross-silo FL,  the clients may not contribute enough data for modeling training due to privacy concerns. To address this issue, we propose a general incentive framework where the profit from the global model can be appropriately allocated to clients to incentivize data contribution.  We formulate the interactions among clients as a data contribution game and analyze its equilibrium. We characterize conditions under which an equilibrium exists, and prove that the client's equilibrium data contribution increases in its data quality but it decreases in its privacy sensitivity. We further conduct numerical experiments using CIFAR-10 and show that the results are consistent with our analysis. Moreover, we find that practical allocation mechanisms such as linearly proportional, leave-one-out, and Shapley-value can incentivize more data contribution from clients with higher-quality data, in which leave-one-out tends to generate the highest global model accuracy.

%\begin{IEEEbiography} [{\includegraphics[width=1in,height=1.25in,clip,keepaspectratio]{./figure/Chao-photo}}]
%{Chao Huang} received the Ph.D degree from the Chinese University of Hong Kong in 2021. During July-November 2021, he was a Post-Doctoral Research Fellow with the Department of Management Sciences, City University of Hong Kong. He is now working as a Post-Doctoral Fellow with the Department of Computer Science, University of California, Davis. His recent research interests span the spectrum of distributed machine learning, network economics, and low-carbon systems.

%\end{IEEEbiography}

%\begin{IEEEbiography}
%	[{\includegraphics[width=1in,height=1.25in,clip,keepaspectratio]{./figure/Huanle-photo}}]
%	{Huanle Zhang} received the Ph.D. degree in computer science from the University of California, Davis, USA, in 2020. He was employed as a project officer with the School of Computer Science and Engineering, Nanyang Technological University, Singapore from 2014-2016. He is a postdoc with the Department of Computer Science, University of California, Davis, from 2020. His research interests include IoT systems and applications, wireless networking, and applied machine learning. 
%	\end{IEEEbiography}

\bibliography{ref,../bib/paper}
\bibliographystyle{IEEEtran}

%
% <OR> manually copy in the resultant .bbl file
% set second argument of \begin to the number of references
% (used to reserve space for the reference number labels box)
%\begin{thebibliography}{1}
%
%%\bibitem{IEEEhowto:kopka}
%%H.~Kopka and P.~W. Daly, \emph{A Guide to \LaTeX}, 3rd~ed.\hskip 1em plus
%%  0.5em minus 0.4em\relax Harlow, England: Addison-Wesley, 1999.
%%  
%
%
%\end{thebibliography}

% that's all folks

\end{document}